\documentclass[12pt]{JHEP3}

\usepackage{amsmath}
\usepackage{amsfonts}
\usepackage{eufrak}
\usepackage{amssymb,bbold}
\usepackage{epsfig}
\usepackage{bm}

\def\be{\begin{equation}}
\def\ee{\end{equation}}

\def\a{\alpha}
\def\b{\beta}
\def\n{\nabla}
\def\t{\tau}

\def\v{\nu}
\def\l{\lambda}
\def\m{\mu}
\def\s{\sigma}
\def\mbx{\mathbf{x}}
\def\o{\omega}

\def\pa{\partial}

\def\O{\Omega}

\def\e{\epsilon}
\def\G{\Gamma}
\def\md{\mathcal{D}}
\def\sq{\sqrt{2}}
\def\d{\delta}
\def\mr{\mathfrak{R}}

\def\g{\gamma}

\def\bfI{\mbox{\boldmath $I$}}
\def\6{\partial}
\def\mf{\mathcal{F}}
\def\mbb{\mathbb{R}}
\def\su2s{(SU(2)\ltimes\mathbb{R}^4)\times\mathbb{R}}
\newcommand{\eref}[1]{(\ref{#1})}

\newcommand{\ip}{\raise1pt\hbox{\large$\lrcorner$}\,}
\newcommand{\bea}{\begin{eqnarray}}
\newcommand{\eea}{\end{eqnarray}}
\newcommand{\nn}{\nonumber \\}

\title{Refining G-structure classifications}

\author{
  Ois\'{\i}n A. P. Mac
  Conamhna\thanks{O.A.P.MacConamhna@damtp.cam.ac.uk} \\ DAMTP \\ Centre
  for Mathematical Sciences \\ University of Cambridge \\ Wilberforce
  Road, Cambridge CB3 0WA, UK.}

\abstract{Using G-structure language, a systematic, iterative
  formalism for computing neccessary 
  and sufficient conditions for the existence of $N$ arbitrary
  linearly independent Killing spinors is
  presented. The key organisational tool is the common
  isotropy group of the Killing spinors. The formalism is
  illustrated for configurations in gauged $SU(2)$ supergravity in
  seven dimensions admitting at least one null Killing spinor, and
  the possible isotropy groups are shown to be $\su2s$, $SU(2)$,
  $\mbb^5$, or the identity. The constraints associated with the
  existence of certain additional Killing spinors are computed, and used to
  derive numerous solutions. A discussion of the relevance of the
  formalism to the complete classification of all supersymmetric configurations
  in $d=11$ is given.} 

\preprint{DAMTP-2004-79}
\begin{document}

\section{Introduction}
In recent years, it has been realised that the notion of a G-structure
is a powerful tool in classifying supersymmetric geometries in
supergravity theories. It has been succesfully and fruitfully applied
in numerous interesting contexts, for example \cite{tod}-\cite{us}. The strategy
is simple. One assumes the existence of at least one Killing
spinor. The existence of a Killing spinor is equivalent to the
existence of a set of globally defined bilinears,
specifying the G-structure, which are invariant under the isotropy group
$G$ of the spinor. Applying the Fierz identities to the bilinears allows
the deduction of algebraic relations between them. Next applying the
Killing spinor equation to the bilinears determines the intrinsic
torsion of the G-structure in terms of the bosonic fields of the
theory. If the theory contains any additional fermions, the vanishing
of their supersymmetry variations implies additional algebraic
relationships between the bosonic fields. When one has derived the
complete set of constraints, one substitutes them back into the
supersymmetry transformations and shows that they are also sufficient
for supersymmetry. One thus arrives at a set of necessary and
sufficient conditions for a bosonic configuration to admit at least
one Killing spinor. However the existence of at least one Killing
spinor generically implies that some but not all of the equations of
motion and Bianchi identities are identically satisfied, so some
subset thereof must still be imposed on the most general field
configuration consistent with the constraints. The standard
G-structure scheme should thus be thought of as providing the most
general supersymmetric ansatz for the supergravity in question.

The standard G-structures formalism has proven to be extremely powerful, and useful from the
point of view of explicitly constructing new solutions, in simple
lower dimensional supergravities with few supercharges. In these
simple cases the computational effort required to arrive at the
general ansatz is also fairly minimal. However as the dimensionality
of spacetime
and/ or the number of supercharges increases, the constraints implied
by the existence of a single Killing spinor become proportionately
weaker, and the most general ansatz becomes broader. Though the
reduced problem of solving the remaining field equations for the most
general supersymmetric ansatz is very much simpler than trying to solve the original
full set {\it ab initio}, it is generically hard to do. For example,
while it is conceptually beautiful that the most general
supersymmetric ansatz for $d=11$ supergravity may be explicitly
computed \cite{gaunt1}, \cite{gaunt3}, the ansatz is not enormously
useful, given current techniques, when it comes to
explicitly generating new solutions. Furthermore, calculating the ansatz
for more complicated theories using the standard approach requires a
lot of (in fact, largely redundant) computation. For a theory such as
IIB, the ammount of calculation required to derive the ansatz using
the standard procedure would be
truly gargantuan. 

There are thus two related improvements to the
standard procedure which are required to fully realise the power the
G-structure formalism. The first, of a purely technical nature, is to
give a more efficient means of performing the calculations. The second
is to give a systematic formalism for classifying configurations
preserving more than one supersymmetry. Some attempts in this
direction have been made, for simple supergravities, using a mixture of G-structure and
integrability techniques \cite{gaunt}, \cite{caccia}, \cite{us}. However this approach will not really be viable for
more complicated theories; it would be preferable to have a
universally applicable formalism
which uses G-structure language throughout, as was done (in a
purely Riemannian context) in \cite{Waldram}. Requiring a configuration to
admit more than one Killing spinor would imply more constraints on the
bosonic fields of the theory. Furthermore it would also imply that
more of the field equations and Bianchi identities would be
identically satisfied. Thus a refined G-structure classification
scheme would be much more useful from the point of view of explicitly
constructing supersymmetric solutions. The objective of this paper is
to propose a means of implementing these improvements to the standard
procedure, illustrated in the context of $SU(2)$ gauged supergravity
in seven dimensions, a theory with sixteen real supercharges.  

The technical novelty employed, which in fact renders a refined
classification tractable, is to fully exploit throughout the
calculation a point raised in \cite{gaunt3}. The point is that any
spinor defines a preferred orthonormal basis of spacetime, and in this
basis the associated G-structure simplifies dramatically, and the
spinor is in fact constant. For example, in the seven dimensional
context of this paper, we will see below that a single null spinor
defines an $(SU(2)\ltimes\mathbb{R}^4)\times\mathbb{R}$ structure. A
single pair of null symplectic majorana spinors (we will work with
symplectic majorana spinors throughout) may be fixed by the projections
\bea
\G^{12}\e^1&=&i\e^2,\nonumber\\\G^{13}\e^1&=&-\e^2,\nonumber\\\G^{14}\e^1&=&i\e^1,\nonumber\\\label{projj}\G^5\e^a&=&\e^a,
\eea
in the basis
\be\label{met}
ds^2=-2e^+e^-+\d_{ij}e^ie^j+(e^5)^2.
\end{equation} 
The four dimensional Riemannian manifold with metric $\d_{ij}e^ie^j$
will be referred to as the base. The various bilinears may be trivially computed without having to
invoke the full Fierzing machinery, and have constant components in
this spacetime basis. This is what was done in \cite{gaunt3}. 

We will take
this simplification a stage further. The supergravity we analyse has
two fermions, $\l$ and $\psi_{\m}$. Consider first the constraints
implied by the vanishing of $\d\l$. The standard way of analysing
these would be to contract $\d\l=0$ with all spinors of the form
$\overline{\e}\G^{(n)}$ up to three gamma matrices to form all
possible combinations of bilinears, and deduce the implied constraints
on the bosonic fields. However it is much more efficient to first
impose the projections (\ref{projj}). Then as we shall see below,
$\d\l=0$ may be cast in the schematic form
\be\label{form}
\d\l=(q+iq^AT^A+q^i\G_i+(r+ir^{A}T^A)\G^-+r^{i}\G^-\G_i)\e=0,
\end{equation}
where the $T^A$ are the Pauli matrices, and we have suppressed
symplectic Majorana indices. In doing this, we shall see below that we
are rewriting $\d\l$
manifestly in terms of a basis for spinor space; by linear
independence, each of the $q$, $q^A$, $q^i$, $r$, $r^A$, $r^i$ must
vanish separately. By imposing the
projections and decomposing in this fashion, it is very much quicker
to derive the constraints. Next consider the Killing spinor
equation. The standard procedure for analysing the differential
constraints would be to apply the full Killing spinor equation to each
of the bilinears to deduce the various components of the spin
connection. Again, this is not efficient. In its preferred basis,
the spinor $\e$ has constant components, so in this basis the Killing spinor
equation becomes schematically
\be
\d\psi_{\m}=\frac{1}{4}\o_{\m\v\s}\G^{\v\s}\e+fluxes=0,
\end{equation}
and thus yields algebraic relations between the fluxes and the
spin connection in the preferred basis. Again we may impose the
various defining projections on $\e$ to reduce each spacetime
component of the Killing spinor equation to the schematic form
\be\label{pap}
\d\psi_{\m}=(q_{\m}+iq^A_{\m}T^A+q^i_{\m}\G_i+(r_{\m}+ir^A_{\m}T^A)\G^-+r^i_{\m}\G^-\G_i)\e=0. 
\end{equation}
Requiring by linear independence that each term in each spacetime
component of the Killing spinor equation vanishes separately allows
one to fix the spin connection in terms of the fluxes, and thus
deduce the most general supersymmetric ansatz with a
minimum of effort. This streamlined way of computing the constraints
has also recently been advocated in \cite{minasian}. 

Having sketched the technical innovation used to reduce as much as
possible the ammount of computation required for the broadest ansatz,
we now turn to the question of refining the classification. We will assume
the existence of the single null Killing spinor defined by the projections
(\ref{projj}), and wish to compute the further constraints on the
bosonic fields of the theory for it to admit an arbitrary additional
linearly independent Killing spinor. An important observation in
organising the refined classification is the following. Incorporating
additional Killing spinors can have one of two effects. Either the
existence of an additional Killing spinor implies a further global
reduction of the structure group of the frame bundle (so the structure
group is reduced from $G$ to some subgroup), or it implies additional
restrictions on the intrinsic torsion of the existing G-structure. To
illustrate this point, consider
the basis of sixteen spinors defined by the projections (with no sum
on $i$)
\bea
\G^{12}\e^1_{(i)}&=&i\a^1_{(i)}\e^2_{(i)},\nonumber\\\G^{13}\e^1_{(i)}&=&-\a^2_{(i)}\e^2_{(i)},\nonumber\\\G^{14}\e^1_{(i)}&=&i\a_{(i)}^3\e^1_{(i)},\nonumber\\\label{proj}\G^5\e^a_{(i)}&=&\a^4_{(i)}\e^a_{(i)},
\eea
for all sixteen possible combinations of $\a^1,...,\a^4=\pm1$. All
these spinors are null, and they are all constant in the spacetime
basis (\ref{met}). We denote them by $(\a_1,\;\a_2,\;\a_3,\;\a_4)$, and also
introduce the notation
\bea
\prod_{j=1}^4\a^j&=&\gamma,\\\prod_{A=1}^3\a^A&=&\b.
\eea
Our basis spans the space of spinors in the theory, so any
additional Killing spinor, whether timelike or null, must be a linear combination
of these spinors; for an arbitrary additional Killing spinor $\e_K$,
\be\label{lin}
\e_K=\sum_{i=1}^Nf_{(i)}\e_{(i)},
\end{equation}
where the sixteen $f_{(i)}$ are real functions. We may rewrite
our basis spinors in terms of our fiducial Killing spinor $\e=(+,+,+,+,)$;
the three other basis spinors with $\gamma=1$, $\b=1$ may be written
as
\be
iT^A\e,
\end{equation}
since these obey the appropriate projections. Next the four basis
spinors with
$(\gamma,\b)=(+,-)$ are given by
\be
\G^i\e,
\end{equation}
while the two sets of four basis spinors with $(\gamma,\b)=(-,+),(-,-)$ are
\be
(\G^-\e,\;i\G^-T^A\e);\;\;\;\;\G^-\G^i\e,
\end{equation}
respectively. Hence an arbitrary additional linearly independent
Killing spinor is given by
\be\label{Killing}
\e_K=(\d+i\d^AT^A+\d_i\G^i+\G^-(\theta+i\theta^AT^A+\theta_i\G^i))\e,
\end{equation}
and we recognise in equations (\ref{form}) and (\ref{pap}) precisely
the decomposition of the supersymmetry variations on the basis. Now,
we will compute the common isotropy group of the spinor $\e_K$ and the
fiducial Killing spinor $\e$ for various choices of the sixteen
functions specifying $\e_K$. We note that basis spinors with $\g=1$
are annihilated by $\G^+$:
\be
\G^+\e_{(i)}=0,\;\;\g_{(i)}=1,
\end{equation}
while basis spinors with $\g=-1$ are annihilated by $\G^-$,
\be
\G^-\e_{(i)}=0,\;\;\g_{(i)}=-1.
\end{equation}
Furthermore, since $\b$ labels minus the chirality of the basis spinors on
the four dimensional base, anti selfdual linear combinations of the $\G^{ij}$
annihilate basis spinors with $\b=1$; for a two form $A^{(-)}_{ij}$
which is anti selfdual on the base but otherwise arbitrary,
\be
A^{(-)}_{ij}\G^{ij}\e_{(i)}=0,\;\;\b_{(i)}=1.
\end{equation}
Throughout this work, for any object with two antisymmetric indices
$i$, $j$ on the base, the superscripts $(+)$ and $(-)$ will denote
respectively the
selfdual and antiselfdual projections on $i$, $j$. Our choice of
orientation, together with all other conventions, is given in Appendix
A. Next, basis spinors with $\b=-1$ are annihilated by selfdual linear
combinations; for an arbitrary selfdual form $A^{(+)}_{ij}$,
\be
A^{(+)}_{ij}\G^{ij}\e_{(i)}=0,\;\;\b_{(i)}=-1.
\end{equation}
Therefore, the most general element of the Lie algebra of $Spin(1,6)$
which annihilates the fiducial Killing spinor $\e$ is
\be\label{isot}
B^AK^{(-)A}_{ij}\G^{ij}+B_i\G^{+i}+B\G^{+5},
\end{equation}
which generates the group
$(SU(2)\ltimes\mathbb{R}^4)\times\mathbb{R}$; the $K^A$ are a triplet
of anti selfdual forms obeying the quaternionic algebra. However, the three
additional Killing spinors with $(\g,\b)=(+,+)$ are also annihilated
precisely by (\ref{isot}); all the spinors in the four dimensional
subspace spanned by the $(+,+)$ basis spinors share the same isotropy
group. Thus, additional linearly independent Killing spinors of the
form
\be
\e_K=(\d+i\d^AT^A)\e
\end{equation}
do not imply any further reduction of the structure group. However,
they will imply further restrictions on the intrinsic torsion of the
$(SU(2)\ltimes\mathbb{R}^4)\times\mathbb{R}$ structure. Since none of
the other twelve basis spinors are annihilated by (\ref{isot}), we see that
an $(SU(2)\ltimes\mathbb{R}^4)\times\mathbb{R}$ structure is
compatible with having one, two, three or four linearly independent
Killing spinors. 

Next consider Killing spinors which are linear combinations of $(+,+)$
and $(-,+)$ spinors, 
\be\label{su2}
\e_K=(\d+i\d^AT^A+\G^-(\theta+i\theta^AT^A))\e,
\end{equation}
 with at
least one of the $\theta$, $\theta^A\neq0$. An arbitrary $(-,+)$ spinor is annihilated by a different
$(SU(2)\ltimes\mathbb{R}^4)\times\mathbb{R}\subset Spin(1,6)$; the
most general Lie algebra element annihilating such a spinor is
\be
B^AK^{(-)A}_{ij}\G^{ij}+B_i\G^{-i}+B\G^{-5}.
\end{equation}
Thus in this case the common isotropy group of the spinors $\e$,
$\e_K$ is $SU(2)$, generated by
\be
B^AK^{(-)A}_{ij}\G^{ij},
\end{equation}
and so additional Killing spinors of the form (\ref{su2}) reduce the structure sroup
to $SU(2)$. There are eight linearly independent spinors with this
common isotropy group, so given the existence of the fiducial Killing spinor,
an $SU(2)$ structure is compatible with the existence of 2,3,...,8
linearly independent Killing spinors. The more Killing spinors there
are, the more restrictive the constraints on the torsion will be.

It is clear how to proceed. Next consider a Killing spinor of the form
\be
\e_K=(\d+i\d^AT^A+\d_i\G^i)\e,
\end{equation}
with at least one of the $\d_i\neq0$. The
$(SU(2)\ltimes\mathbb{R}^4)\times\mathbb{R}$ isotropy group of a
$(+,-)$ spinor is generated by 
\be
B^AJ^{(+)A}_{ij}\G^{ij}+B_i\G^{+i}+B\G^{+5},
\end{equation}
so in this case the common isotropy group of the spinors $\e$, $\e_K$
is $\mathbb{R}^5$, generated by
\be
B_i\G^{+i}+B\G^{+5}.
\end{equation}
As in the $SU(2)$ case, there are eight linearly independent spinors with
this common isotropy group, so given the existence of the fiducial
Killing spinor, an $\mathbb{R}^5$ structure is compatible with the
existence of 2,...,8 linearly independent Killing spinors.

It is easy to verify that assuming the existence of Killing spinors
which are more general linear combinations of the basis spinors than
the three cases discussed above imply that the structure group is
reduced to the identity. Such a structure, given the existence of the
fiducial Killing spinor, is compatible with the existence of
2,3,...,16 linearly independent Killing spinors. 

In the refinement of the G-structure classification scheme 
proposed here, configurations admitting multiple linearly independent
Killing spinors are naturally classified according to the structure
group, rather than the number of Killing spinors. We have seen how we
can have $G=(SU(2)\ltimes\mathbb{R}^4)\times\mathbb{R}$ and four
linearly independent Killing spinors, or $G=Id$ and only two. This is
somewhat different to other approaches, for example generalised holonomy
\cite{duff}-\cite{tsimpis}, where the aim has always been
to classify configurations according to the number of preserved
supersymmetries. From our perspective, this is rather unnatural; the
most significant feature of a supersymmetric configuration in the
G-structure formalism is the structure group. Then for a given
structure group, demanding more supersymmetries imposes progressively
more severe constraints on the intrinsic torsion.  

The most efficient way that we have been able to find of computing the
additional constraints implying and implied 
by the existence of arbitrary 
additional Killing spinors is the following. Denote the supersymmetry
variations  $\d\l$, $\d\psi$ with parameter
$\e_K$ as $\Delta_{\l}\e_K$, $\mathcal{D}_{\m}\e_K$. Since we may write
$\e_K=Q\e$, for some $Q$ of the form of equation (\ref{Killing}),  we
observe that since $\e$ is Killing, $\e_K$ is Killing if and only if
\bea\label{ptooop}
\left[ \Delta_{\l},Q \right] \e&=&0,\\\label{ptoooop}\left[ \mathcal{D}_{\m},Q \right] \e&=&0.
\eea
Having evaluated the commutators, we may as before impose the
projections satisfied by $\e$ to reduce these expressions to manifest linear
combinations of the basis spinors, and by linear independence, each
coefficient must vanish separately. The procedure may be iterated at
will, for any desired choices of $Q$, consistent with any desired
structure group. For simple supergravities, such as those with eight
supercharges, the procedure given here should be easy to employ to perform a
fully refined classification, and should involve a modest amount
of calculation. For a theory of the complexity of the one studied in
this paper, with sixteen supercharges, the amount of computation
required to perform a complete classification is much
larger. This is because the most general additional linearly
independent Killing spinor consistent with the existence of an
identity structure is parameterised by sixteen real functions, and
keeping track of all the terms in (\ref{ptooop}), (\ref{ptoooop}) is 
technically involved. Additional Killing spinors consistent
with the larger structure groups $\su2s$, $SU(2)$ and $\mbb^5$ are
somewhat easier to handle, since these are parameterised by four, eight and
eight real functions respectively. Nevertheless, we have not performed the refined classification in
its entirety; for
the smaller structure groups, we have restricted to some illustrative
examples, rather than pursuing the problem in full
generality. However, we emphasise that there is no conceptual
difficulty in doing so; the calculations involved, while lengthy, are
simple and repetitive, involving nothing more than evaluating
gamma-matrix commutators and imposing a fixed set of spinor
projections.  
\\
\\
The traditional approach to finding supersymmetric solutions of
supergravity theories has been to make some ansatz for the bosonic
fields, and then to use the supersymmetry variations of the fermions
to determine if there are any Killing spinors consistent with that
ansatz. The G-structures programme can be thought of as the exact
converse of this; one makes an ansatz for the Killing spinors, and
expresses the conditions for a spinorial solution of the supersymmetry
variations of that form to exist as a set of constraints on the
bosonic fields. The power of the formalism lies in the fact that the
spinor ansatz can, if desired, be made completely general, whereby we
mean that the constraints on the bosonic fields for $N$ arbitrary
linearly independent Killing spinors (for any desired structure group)
to exist can be evaluated, albeit
with some effort, in any theory. What we have argued above is that the
formidably complicated problem of determining the constraints on the
bosonic fields for the general multi-spinor solution of the fermion
supersymmetry transformations to exist may naturally be organised into
more manageable subproblems, with G-structures providing the
organising principle. One may classify the possible types of
multi-spinor ans\"{a}tze according to the common isotropy
group of the spinors. Then within each class one may
deduce the constraints on the intrinsic torsion of the structure
implied by the existence of any desired number of arbitrary linearly
independent spinors within that class.
\\
\\  
The remainder of this paper is organised as follows. In section two we
briefly describe $SU(2)$ gauged supergravity in seven dimensions. In
section three  we
perform a streamlined analysis of the constraints implying and implied by the
existence of a single null Killing spinor, and introduce coordinates
for the problem. In section four, with a modest assumption of the form
of the Yang-Mills fields of the theory introduced for computational
convenience, we perform a complete classification of multi-supersymmetric
configurations admitting a strictly $\su2s$ structure. It is shown that
a second Killing spinor with the same structure group
as the first exists if and only if the Yang-Mills fields are truncated
to a $U(1)$ subgroup. The existence of a third Killing spinor with the
same isotropy group implies and is implied by the vanishing of the
Yang-Mills fields, and implies the existence of a fourth Killing
spinor. In section five we study configurations with an $SU(2)$
structure; neccessary and sufficient conditions for the existence of
an additional Killing spinor of a specific form are derived, and also
all configurations admitting eight Killing spinors fixed by the same
$SU(2)$ are explicitly classified. We illustrate the effect
incorporating additional Killing spinors has on the torsion of the
$SU(2)$ structure by deriving known $AdS_3$ solutions of the theory
from our ansatz, and generalise these solutions to construct
(singular) membrane solutions with $AdS_3$
worldvolumes. Configurations with $\mbb^5$ structures are the topic of
section six. In section
seven we discuss identity structures, providing explicit
examples. Section seven contains our conclusions, and a discussion of
the applicability of the formalism to the complete classification of
supersymmetric configurations in eleven dimensions. In appendix A we give our conventions, and appendix B
lists a set of projections satisfied by $\e$ which we use
throughout. In appendix C we give the components of the spin
connection. Appendix D contains the integrability conditions for the
theory. 

\section{The supergravity}
Minimal $SU(2)$ gauged supergravity in seven dimensions was first
constructed in \cite{Townsend:1983kk} but with numerical typos, which
were corrected in \cite{Mezincescu:ta}. In our conventions, the
bosonic lagrangian density for the theory is
\bea
e^{-1}\mathcal{L} & = & \frac{1}{2}  R - \frac{1}{24 } 
(G_{\mu\nu\rho\tau})^2  - \frac{1}{2}   F_{\mu\nu}\,_{\;\;b}^a
F^{\mu\nu }\,^b_{\;\;a} - \frac{5}{2}(\6_\mu \phi)^2 +\frac{h}{18}e^{-2\phi}\e^{\m\v\s\t\a\b\gamma}G_{\m\v\s\t}A_{\a\b\gamma}  \nn &-&
  \frac{1}{24}e^{-\phi}
G_{\mu\nu\rho\tau} F_{\kappa \lambda }\,^a_{\;\;b} A_{\chi }\,^b_{\;\;a}
\e^{\mu\nu\rho\tau\kappa \lambda\chi} -V(\phi). 
\eea
Compared to \cite{Townsend:1983kk} we use the same conventions for the
Riemann tensor but Hawking and Ellis conventions for the Ricci tensor
and scalar.  We have also rescaled $\phi\rightarrow {\sqrt 5}\, \phi$
and the forms by $F\rightarrow {\sqrt 2}\,e^{\phi} F$, $G\rightarrow
{\sqrt 2}\,e^{-2\phi} G$, $A_{(1),(3)}\rightarrow\sq A_{(1),(3)}$. The potential is given by
\be
V(\phi)=-60m^2+10(m^{\prime})^2,
\end{equation}
where $m$ is a function of the single scalar field $\phi$,
\be
m=-\frac{2}{5}he^{-4\phi}-\frac{1}{10}ge^{\phi},
\end{equation}
with $g$ the gauge coupling (we have rescaled the coupling in
\cite{Townsend:1983kk} by $g\rightarrow g/\sq$)
and $h$ the (constant) topological
mass.
The supersymmetry
variations of the fermions are given by
\bea
    \delta \lambda^a & = & \frac{\sqrt{5}}{2} \G^\mu D_\mu \phi \, \e^a +
    \frac{i}{2\sqrt{5}}  \G^{\mu\nu} F_{\mu\nu }\,^a_{\;\;b} \, \e^b +
    \frac{1}{24\sqrt{5}} 
    \G^{\mu\nu\rho\tau} G_{\mu\nu\rho\tau} \, \e^a -\sqrt{5}m^{\prime}\e^a,  \\ 
    \delta \psi_{\mu }^a & = & D_\mu \, \e^a - \frac{i}{10} 
    (\G_\mu\,^{\nu\rho}-8 \delta^\nu_\mu \G^\rho) F_{\nu\rho }\,^a_{\;\;b} \,
    \e^b + \frac{1}{80} 
    (\G_\mu\,^{\alpha\beta\gamma\delta}-\frac{8}{3}
    \delta^\alpha_\mu \G^{\beta\gamma\delta})
    G_{\alpha\beta\gamma\delta}\, \e^a \nn
    &+&m\G_{\m}\e^a-igA^{\;\;a}_{\m b}\e^b , \label{eq:Killing} 
\eea
and the parameter $\e^a$ is a symplectic-Majorana spinor, whose
properties are summarized in appendix A.  

Let us introduce the following notation. Let $A_p$, $B_q$ be $p$-
and $q$-forms respectively. Then  
\be
A\lrcorner
B_{a_1...a_{q-p}}=\frac{1}{p!}A^{b_1...b_p}B_{b_1...b_pa_1...a_{q-p}}.
\end{equation}
The equations of motion and Bianchi identities are
\begin{eqnarray}
   d(e^{-2\phi}G) &=& 0, \\ 
   d(e^{\phi}F^A) &=& g\e^{ABC}e^{\phi}F^B\wedge A^C,\\  
   P &=& 5\n^2\phi-4G\lrcorner G+F^A\lrcorner F^A-V^{\prime}=0, \label{eq:P} \\ 
   Q &=& \star(e^{-2\phi}d\star(e^{2\phi}G)-\frac{1}{2}F^A\wedge
   F^A+8he^{-4\phi}G)=0, \label{eq:Q} \\ 
   R^A &=& \star(e^{\phi}d\star(e^{-\phi}F^A)-g\e^{ABC}e^{\phi}\star
   F^B\wedge A^C-2F^A\wedge G)=0,
   \label{eq:R} \\
   E_{\m\v}&=&R_{\m\v}-\frac{1}{3}\Big(G_{\m\a\b\gamma}G_{\v}^{\;\;\;\a\b\gamma}-
   \frac{1}{10}g_{\m\v}G_{\a\b\gamma\d}G^{\a\b\gamma\delta}\Big)-5\pa_{\m}\phi
   \pa_{\v}\phi \nonumber\\
   &-& \left(
   F^{A}_{\m\a}F^{A\a}_{\v}-\frac{1}{10}g_{\m\v}F^A_{\a\b}F^{A\a\b}
   \right)-\frac{2}{5}g_{\m\v}V  = 0.
\eea
When $h=0$, the theory lifts on an $S^3$ to the NS sector in
$d=10$ \cite{chamseddine}. When $hg>0$, it lifts on an $S^4$ to $d=11$
\cite{pope}. When $h\neq 0$, there is a subtlety in imposing the
four-form field equation. The reason is that the 3-form $A_{(3)}$,
which is massive, would have twenty on-shell degrees of freedom if it
satisfied an ordinary second order field equation. However the 3-form
in the 7d supergravity multiplet should have only ten on-shell degrees
of freedom. This is achieved by imposing the odd-dimensional
selfduality equation \cite{Pilch}:
\be\label{selfd}
e^{2\phi}\star G-\frac{1}{2}(e^{\phi}F^A\wedge
A^A-\frac{g}{6}\e^{ABC}A^A\wedge A^B\wedge A^C)+8hA_{(3)}=0.
\end{equation}
Note that the exterior derivative of this equation is just $\star
Q$. Imposing the Bianchi identity and $Q=0$ fixes $A_{(3)}$ up to an
arbitrary closed three form. The closed three form is then determined
by demanding that $A_{(3)}$ satisfies (\ref{selfd}). In the examples
given below, we will explicitly impose the Bianchi identity and $Q=0$,
but leave the determination of the closed three form in $A_{(3)}$
implicit.

\section{Configurations admitting at least one null Killing spinor}
\subsection{The G-structure}
It is instructive to verify, using the standard formalism, that a
single null spinor in seven dimensions defines an $\su2s$
structure. It was shown in \cite{us} that we can define the following spinor bilinears
\bea
f^{(ab)} & = & \bar \e^a \e^b \label{eq:scalars} , \\ 
\e^{ab} V_\mu & = &  \bar \e^a \G_\mu \e^b ,  \\
\e^{ab} I_{\mu\nu} & = &  \bar \e^a \G_{\mu\nu} \e^b , \\
\Omega^{(ab)}_{\mu\nu\rho} & = &  \bar \e^a \G_{\mu\nu\rho} \e^b 
\,. \label{eq:3_form} 
\eea
From the reality properties of the gamma matrices and the symplectic
Majorana condition, the vector $V_\mu$ and the two-form  $I_{\mu\nu}$ are seen to be real, while instead the scalars
and the $3$--form can be rewritten as 
\begin{eqnarray} 
     f^a_{\;\; b} &=& - i g^A \, (T^A)^a_{\;\; b} , \\ 
     \Omega^a_{\;\; b} &=& - i X^A \, (T^A)^a_{\;\; b} , 
\end{eqnarray} 
with $g^A$, $X^A_{\mu\nu\rho}$, $A=1,2,3$, real. $(T^A)^a_{\;\; b}=
1/2 (\sigma^A)^a_{\;\; b}$ are generators of the $SU(2)$ Lie algebra,
$\sigma^A$ being the Pauli matrices, and obey 
\be 
    (T^A)^a_{\;\; b}\, (T^B)^b_{\;\; c} = \frac{1}{4}
\delta^{AB}\,\delta^a_c + \frac{i}{2} \epsilon^{ABC}\,(T^C)^a_{\;\; c}
. 
\ee  

One important consequence of the Fierz identity \eref{eq:Fierz_id}
is that $V_\mu$ is either timelike or null 
\be 
     V^2 = -\frac{1}{4} g^A\, g^A . 
\ee 
The case of a single timelike spinor has been analysed in a previous
work \cite{us}. Here we want to identify
the G-structure associated a single null spinor, a spinor such that $g^A=0$. From the Fierz identities we may
deduce
\bea
i_VI&=&0,\\I\lrcorner I&=&0\\I_{\m\t}I^{\v\t}&=&V_{\m}V^{\v}.
\eea
These imply that
\be
I=V\wedge K,\;\;\; K^2=1,\;\;\;.
\end{equation}
Furthermore from the Fierz identities we may deduce the projection
\be\label{projh}
K_{\m}\G^{\m}\e^a=\e^a,
\end{equation}
together with
\be
i_VX^A=i_KX^A=0.
\end{equation}
Let us introduce a null orthonormal basis where
\bea
V&=&-e^+, \;\;K=e^5,\\ds^2&=&-2e^+e^-+\d_{ij}e^ie^j+(e^5)^2,
\eea
$i,j=1,..,4$, and we choose an orientation 
\be
\e^{+-12345}=1.
\end{equation} 
Then from the seven dimensional duality relation for
the gamma matrices, (\ref{projh}) becomes a chirality projection in the
six Lorentzian dimensions orthogonal to $K$. It is well known
\cite{bryant} that a pair of symplectic Majorana-Weyl spinors (or a single
Weyl spinor) defines an $SU(2)\ltimes\mathbb{R}^4$ structure in six
Lorentzian dimensions, so our seven dimensional structure is simply
$(SU(2)\ltimes\mathbb{R}^4)\times\mathbb{R}$. Now from the results of \cite{reall}, we may immediately
read off the form of the $X^A$:
\bea
X^A&=&2e^+\wedge
J^A,\\J^{Ai}_{\;\;\;\;j}J^{Bj}_{\;\;\;\;k}&=&\e^{ABC}J^{Ci}_{\;\;\;\;k}-\d^{AB}\d^{i}_k.
\eea
With our choice of conventions, the $J^A$ are selfdual on the base with orientation
\be
\e^{ijkl}=\e^{+-ijkl5}
\end{equation}
The single spinor obeys the projection (\ref{projh}) and
\be
J^A_{ij}\G^{ij}\e^a=8iT^{Aa}_{\;\;\;\;b}\e^b.
\end{equation}
We may completely fix the spinor by requiring that in addition to
(\ref{projh}) it obeys the
projections 
\bea
\G^{12}\e^1&=&i\e^2,\nonumber\\\G^{13}\e^1&=&-\e^2,\nonumber\\\G^{14}\e^1&=&i\e^1,
\eea
so that the $J^A$ take the canonical form
\be
J^1=e^{12}+e^{34},\;\;\;J^2=-e^{13}+e^{24},\;\;\;J^3=e^{14}+e^{23}.
\end{equation}
For performing the calculations of the
constraints it is very useful to have a set of
projections satisfied by $\e$ which are implied by the defining
projections. We give such a set in appendix B.

\subsection{Constraints for supersymmetry}
Now we solve the constraints
$\d\lambda=\d\psi_{\m}=0$ for the single spinor $\e$. In
order to reduce as much as possible the computation required we will
work with $\star G$, the dual of the four form $G$.
Hopefully without risk of confusion, we also denote the dual by
$G$. Where ambiguity can arise we will explicitly indicate the rank of
the form. In terms of the dual $G$, the supersymmetry variations of the
fermions are 
\bea
\sqrt{5}\d\l^a&=&\frac{5}{2}\pa_{\m}\phi\G^{\m}\e^a+\frac{i}{2}F^A_{\m\v}\G^{\m\v}T^{Aa}_{\;\;\;\;\;b}\e^b-\frac{1}{6}G_{\m\v\s}\G^{\m\v\s}\e^a-5m^{\prime}\e^a,\\\d\psi_{\m}^a&=&\n_{\m}\e^a-\frac{i}{10}(\G_{\m}^{\;\;\;\v\s}-8\d_{\m}^{\v}\G^{\s})F^A_{\v\s}T^{Aa}_{\;\;\;\;\;b}\e^b-\frac{3}{20}G_{\m\v\s}\G^{\v\s}\e^a+\frac{1}{30}G_{\v\s\t}\G_{\m}^{\;\;\;\v\s\t}\e^a\nonumber\\&-&igA^A_{\m}T^{Aa}_{\;\;\;\;\;\b}\e^b+m\G_{\m}\e^a.
\eea

\subsubsection{Constraints from $\d\lambda=0$}
Consider first the supersymmetry variation of $\d\l$. We will in fact use
two shortcuts from a previous work \cite{us}. The
first is that $i_V\pa\phi=0$. Also, on the assumption that $\e$ is
Killing, we have that $i_VF^A=0$. Thus
\bea
\pa_-\phi&=&0,
\\F^A_{-\m}&=&0.
\eea
An implication of
these is that the only place terms with a single $\G^-$ can arise
in $\d\l$ is in $G_{\m\v\s}\Gamma^{\m\v\s}$. Also, all terms with a
single $\G^+$ drop out, since $\G^+\e=0$. Consider the $\G^-$ terms. These
may easily be computed to be 
\be
-\G^-[G_{-i5}\G^i+iG_{-ij}J^{Aij}T^A]\e.
\end{equation}
By linear independence, each of these must vanish
separately. Hence
\bea
G_{-i5}&=&0,\\G_{-ij}^{(+)}&=&0.
\eea
Now all remaining terms in $\d\l=0$ may be converted into terms
containing no gamma matrices or a single gamma matrix on the
base. The terms with a single gamma matrix are given by 
\be
\Big[\frac{5}{2}\pa_i\phi+G_{+-i}+\frac{1}{6}\e_i^{\;\;\;jkl}G_{jkl}+\frac{1}{2}F^A_{j5}J^{Aj}_{\;\;\;\;\;i}\Big]\G^i\e,
\end{equation}
and each component of the one form in square brackets must
vanish. Finally, the terms 
involving no gamma matrices are of the form $(A+i\sum_AB^AT^A)\e$, and
by linear independence we must have
$A=B^A=0$. This implies that
\bea
G_{+-5}+\frac{5}{2}\pa_5\phi-5(m+2he^{-4\phi})&=&\frac{1}{4}F^A_{ij}J^{Aij},\\G^{(+)}_{5ij}&=&-\frac{1}{8}\e^{ABC}F^A_{kl}J^{Bkl}J^C_{ij}.
\eea
This completes the analysis of $\d\l^{(i)}=0$.

\subsubsection{$\d\psi_{\m}=0$}
Next we turn to the analysis of the Killing spinor equation. We will
use another result from \cite{us}, namely
that for null Killing spinors,
\be
\n V=i_VG-d\phi\wedge V-4he^{-4\phi}I,
\end{equation}
so that $V$ is a Killing vector and
\bea
\o_{\m\v-}&=&G_{\m\v-}+(d\phi\wedge
e^+)_{\m\v}+4he^{-4\phi}(e^+\wedge e^5)_{\m\v}.
\eea
Also we wish to comment on the choice of gauge. We always can, and
will, choose the gauge $i_VA^A=0$. We also know that
$i_VF^A=0$. Therefore if $v$ is the coordinate along the integral
curves of $V$, we have that $A^A_{\m}$ is
independent of $v$, in the gauge $i_VA^A=0$. Therefore since we still have the freedom to
perform $v$-independent gauge transformations, we may always choose a
gauge such that 
\be
A^A_+=A^A_-=0,
\end{equation}
and this is the gauge we work in henceforth. Now we know that each
spacetime component of the
Killing spinor equation can be reduced to the form (\ref{pap}) given
in the introduction. All terms involving a single $\G^+$ drop
out. Furthermore, we have explicitly checked that all the terms of the
form
\be
(q_{\m}+\G^-(r_{\m}+ir_{\m}^AT^A+r_{\m i}\G^i))\e
\end{equation}
vanish identically as a consequence of the constraints we have already
derived. Thus the Killing spinor equation reduces to
\be\label{pakpak}
\d\psi_{\m}=(iq_{\m}^AT^A+q_{\m i}\G^i)\e.
\end{equation}
Using the projections of appendix B, it is easy to see that the spin
connection components $\o_{\m i5}$ contribute only to the $q_{\m
  i}\G^i\e$ terms, the $\o_{\m ij}^{(-)}$ components drop out, and the
$\o_{\m ij}^{(+)}$ components contribute only to the $iq^A_{\m}T^A\e$
terms. The surviving components of the spin connection may be fixed in
terms of the fluxes by demanding that the linearly independent terms
(\ref{pakpak}) vanish.  Thus, from the + component of $\d\psi_{\m}=0$ we find  
\bea
\o_{+ij}^{(+)}&=&G_{+ij}^{(+)}-\frac{1}{2}F^A_{+5}J^A_{ij},\\\o_{+i5}&=&G_{+i5}-F^A_{+j}J^{Aj}_{\;\;\;\;\;i}.
\eea
From the $-$ component we deduce
\bea
\o_{-ij}^{(+)}&=&0,\\\o_{-i5}&=&0.
\eea
The 5 component gives
\bea
\o_{5ij}^{(+)}&=&G^{(+)}_{5ij}+\frac{g}{2}A^A_5J^A_{ij},\\
\o_{55i}&=&4\pa_i\phi+2G_{+-i}+\frac{1}{3}\e_{ijkl}G^{jkl}.
\eea
Finally from the $i$ component we get
\bea
\o_{ij5}&=&(
G_{+-5}+\frac{3}{2}\pa_5\phi-5m-6he^{-4\phi})g_{ij}+G_{5ij}^{(-)}-G_{5ij}^{(+)}\nonumber\\
&-&F^{A(-)}_{ik}J^{Ak}_{\;\;\;\;\;j}\\\o_{ijk}J^{Ajk}&=&2gA^A_i-F^A_{i5}+\e^{ABC}F^B_{j5}J^{Cj}_{\;\;\;\;\;i}\nonumber\\&-&(3\pa_j\phi+G_{+-j})J^{Aj}_{\;\;\;\;\;i}\Big.
\eea
This concludes the analysis of $\d\psi_{\m}=0$. The practical
advantages of calculating the necessary and sufficient conditions for
supersymmetry in this fashion are obvious. From the integrability
conditions of Appendix D, we may deduce that it is sufficient to
impose the Bianchi identity for the Yang-Mills fields and $G_{(4)}$,
the field equation for $G_{(4)}$, the $+$ component of the Yang-Mills
field equation and the $++$ component of the Einstein equation on this
supersymmetric ansatz. All other field equations are satisfied identically.

\subsection{Introducing coordinates}
In this subsection we will introduce coordinates for our problem. Our
null Killing spinor induces a natural coordinate system in which the
associated symmetries of the metric are manifest. Closely following \cite{gaunt3}, we introduce coordinates $(v,u,z,x^M)$ such that
the vectors dual to the basis one forms are
\bea
e^+&=&-\frac{\pa}{\pa v},\\e^-&=&-\frac{L\mathcal{F}}{2}\frac{\pa}{\pa
  v}+L\frac{\pa}{\pa u},\\e^5&=&-\frac{B}{C}\frac{\pa}{\pa
  v}+\frac{1}{C}\frac{\pa}{\pa z},\\e^i&=&e^{i\m}\pa_{\m}.
\eea
Inverting we get
\bea
e^+&=&-L^{-1}(du+\l),\\e^-&=&dv+\frac{1}{2}\mf
du+Bdz+\v,\\e^5&=&C(dz+\s),\\e^i&=&e^i_Mdx^M.
\eea
We have in fact fixed some of the gauge freedom in the metric. Now
repeating an argument given in \cite{gaunt3}, we can deduce that
$L,\:\mf,\:B,\:C,\:\l,\:\v,\:\s$ and $e^i_M$ may all be taken to be
independent of $v$. The components of the spin connection for this
metric were computed in \cite{gaunt3}. We have performed the trivial
modification to adjust to our slightly different conventions, and we
give the result in appendix C, where we use the following
notation. Let $Q$ be a q-form on the base, satisfying $\mathcal{L}_VQ=0$:
\be
Q=\frac{1}{q!}Q(u,z,x^M)_{M_1...M_q}dx^{M_1}\wedge...\wedge dx^{M_q}.
\end{equation}
Define the exterior derivative restricted to the base
\be
\tilde{d}Q\equiv\frac{1}{q!}\frac{\pa}{\pa
  x^{M_1}}Q_{M_2...M_{q+1}}dx^{M_1}\wedge...\wedge dx^{M_{q+1}},
\end{equation}
and denote the Lie derivatives with respect to $\frac{\pa}{\pa u}$ and
$\frac{\pa}{\pa z}$ acting on such forms by $\pa_uQ$, $\pa_z Q$
respectively. Then, defining
\be
\md Q\equiv \tilde{d}Q-\s\wedge\pa_zQ+\l\wedge\pa_uQ,
\end{equation}
we have
\be
dQ=\md Q-Le^+\wedge\pa_uQ+C^{-1}e^5\wedge\pa_zQ.
\end{equation}
Finally we define the following quatities:
\bea
M_{ij}&=&\d_{ik}(\pa_u e^k)_j,\\\Lambda_{ij}&=&\d_{ik}(\pa_ze^k)_j.
\eea
From the expression for the spin connection components in appendix C,
we may rewrite the constraints $\o_{-i5}=\o_{-ij}^{(+)}=0$ as
\bea
\l&=&\l(u,x),\\\md\l^{(+)}&=&0.
\eea

\subsection{Refining the classification}
In this section we have given the full set of neccessary and sufficient
conditions for the existence of a single arbitrary null spinor. In the
next four sections we will illustrate the computation of the
constraints associated with the existence of additional spinors, for
each of the four possibilities for the structure groups. As
discussed in the introduction, this involves computing the expressions
\bea\label{comm}
\left[\Delta_{\l},Q\right]\e&=&0,\\\left[\mathcal{D}_{\m},Q\right]\e&=&0,
\eea
for a choice of $Q$ compatible with the desired G-structure, and
reducing them to canonical form, as a manifest sum of basis
spinors. As has been stated, to do this for the most general choices
of $Q$ requires a
lot of computation, and we will not perform the refined classification
in full generality. For the structure groups $SU(2)$, $\mbb^5$ and the
identity, we will restrict attention to illustrative
examples, for particular choices of $Q$. This means that we will make
specific ans\"{a}tze for the additional Killing spinors. Also for
computational convenience, unless otherwise stated we will restrict attention to bosonic
configurations for which the Yang-Mills fields are of the form
\bea\label{ymform}
A^A&=&A^A_i(x)e^i,\nn
F^A&=&\frac{1}{2}F^A_{ij}(x)e^i\wedge e^j,
\eea
but which are otherwise general. Note that this is the only assumption
we make about the form of the bosonic fields; all the additional
constraints we derive will follow as a consequence of the choice of $Q$.

\section{$(SU(2)\ltimes\mathbb{R}^4)\times\mathbb{R}$ structure}
As we saw in the introduction, the simplest additional Killing spinors
to incorporate are those which share the same isotropy group as the
fiducial Killing spinor, and so imply no further reduction of the
structure group. These spinors are parameterised by four real
functions; given our assumption of the form of the Yang-Mills fields,
we will completely analyse all $\su2s$ structures with more than one
Killing spinor. In the first subsection we will derive the
constraints, and in the second subsection we will construct a class of
solutions for which the structure group is strictly $\su2s$ and not
some subgroup.

\subsection{Constraints} 
Consider an arbitrary $(+,+)$ spinor, which we denote
by
\be
(\d+i\d^AT^A)\e.
\end{equation}
As discussed in the introduction, assuming that this spinor is Killing
does not imply any reduction of the structure group, but it does yield
additional constraints on the intrinsic torsion. The commutators
(\ref{comm}) for such spinors are easy to compute,
since they just pick out the Yang-Mills terms in the supersymmetry
variations. First consider
\be
\left[\Delta_{\lambda},\d+i\d^AT^A\right]\e=-\frac{1}{2}F^A_{ij}\G^{ij}\d^B\left[T^A,T^B\right]\e.
\end{equation}
Reducing to canonical form, we get
\be
\Big(\frac{1}{4}\e^{ABC}F^A_{ij}\d^BJ^{Cij}+\frac{i}{2}(F^B_{ij}J^{Bij}\d^A-F^A_{ij}J^{Bij}\d^B)T^A\Big)\e.
\end{equation}
Next consider the Killing spinor equation. The commutator is
\be
\left[\n_{\m}-i(gA^A_{\m}+\frac{1}{10}(\G_{\m\v\s}-8g_{\m\v}\G_{\s})F^{A\v\s})T^A,\d+i\d^BT^B\right]\e.
\end{equation}
The $-$ component is particularly easy to evaluate. It reduces to 
\be
(\pa_-\d+i\pa_-\d^AT^A)\e,
\end{equation}
The $+$
component is
\be
(\pa_+\d+i\pa_+\d^AT^A+\frac{1}{5}\G^-\left[\Delta_{\lambda},i\d^AT^A\right])\e.
\end{equation}
The 5 component is
\be
(\pa_5\d-\frac{1}{5}\left[\Delta_{\lambda},i\d^AT^A\right]+i\pa_5\d^AT^A)\e,
\end{equation}
and for the $i$ component we have
\bea
&&\Big[\pa_i\d+i(\pa_i\d^A+g\e^{ABC}A^B_i\d^C)T^A)-\frac{1}{2}\e^{ABC}\Big(\frac{3}{5}F^{A(+)}_{ij}+F^{A(-)}_{ij}\Big)\d^BJ^{Cj}_{\;\;\;\;\;k}\G^k\Big]\e.\nonumber
\eea

Now, requiring the vanishing of the commutators acting on $\e$ implies
the 
following algebraic restrictions on the Yang-Mills fields in terms of
the parameters $\d^A$:
\bea
\e^{ABC}F^A_{ij}\d^BJ^{Cij}&=&0,\nn
\d^AF^B_{ij}J^{Bij}-\d^BF^A_{ij}J^{Bij}&=&0,\nn\e^{ABC}F^{A(-)}_{jk}\d^BJ^{Cj}_{\;\;\;\;\;i}&=&0.
\eea
We
also get a set of differential constraints on the $\d,\d^A$ which are
\bea
\pa_{\m}\d&=&0,\\\pa_-\d^A=\pa_+\d^A&=&\pa_5\d^A=0,\\\label{mono}\pa_i\d^A&=&-g\e^{ABC}A^B_i\d^C.
\eea
Since we have already assumed that $\e$ is Killing, and we have
derived that $\d=$const, we take $\d=0$. We
also have
\bea
\d^A&=&\d^A(x).
\eea
Given the Yang-Mills Bianchi identity, the integrability condition for (\ref{mono}) implies that
$\e^{ABC}F^B_{ij}\d^C=0$, and so implies the algebraic
constraints. Therefore a bosonic configuration admitting the Killing spinor
$\e$ also admits the Killing spinor $i\d^AT^A\e$ if and only if
equation (\ref{mono}) is satisfied. By performing an $x$ dependent
$SU(2)$ gauge transformation (thus preserving $A^A_+=A^A_-=A_5^A=0$)
we may set
\be
\d^A=k\d^{1A},
\end{equation}
for some constant $k$ which can be eliminated by a constant rescaling,
and so the existence of this Killing spinor implies and is implied by
the condition that the 
Yang-Mills field is truncated to a $U(1)$ subgroup,
\be
A^2=A^3=0,
\end{equation}
Clearly, requiring the existence of a third linearly independent
$(+,+)$ Killing spinor would imply that
\be
A^1=0,
\end{equation}
and then the configuration automatically admits a fourth linearly
independent spinor. Thus, the
$(SU(2)\ltimes\mathbb{R}^4)\times\mathbb{R}$ structure can admit one,
two, or four linearly independent Killing spinors, depending on
whether the Yang-Mills fields are $SU(2)$, $U(1)$ or
zero. Incorporating additional $(+,+)$ spinors implies no further
constraints on $G$, $\phi$ or $\o$, beyond those derived in 
section 2. From these results, we also see that when the Yang-Mills
fields are truncated to a $U(1)$ subgroup, Killing spinors always come
in pairs; if $\e_K$ is Killing, then so is the linearly independent
spinor $iT^1\e_K$, since $T^1$ commutes with the operators in the fermion supersymmetry
variations. Similarly, when the Yang-Mills fields vanish, Killing
spinors always come in groups of four.

\subsection{Examples}
Let us now present a family of solutions for which the structure
group is strictly $\su2s$ and not some subgroup. It is easy to verify
that the vacuum solution
\be\label{su2sb}
ds^2=(4hz)^{-2}(-dt^2+dr^2+dz^2+ds^2(\mathcal{M}_4)),
\end{equation}
where $\mathcal{M}_4$ is hyperk\"ahler and $16h=g$, satisfies all the
constraints of section 2, and the $++$ component of the Einstein
equations is satisfied. Since the Yang-Mills fields vanish, these
solutions admit Killing spinors in multiples of four. When
$\mathcal{M}_4=\mbb^4$, this is nothing but the maximally
supersymmetric $AdS_7$ solution of the theory. Choosing any other
hyperk\"{a}hler base reduces the number of independent Killing spinors
to four. One may show that demanding the existence of an arbitrary
additional Killing spinor of the form (\ref{Killing}) implies the
constraints 
\bea
\label{delti}\tilde{\n}_i\d_j&=&0,\\\pa_{\m}\theta&=&\pa_{\m}\theta^A=0,\\
\label{deltii}\tilde{\n}_i\theta_j&=&2m\theta \d_{ij}+m\theta^AJ^A_{ij},
\eea
where $\tilde{\n}$ denotes the Levi-Civita connection on the
base. Given that for vanishing Yang-Mills fields Killing spinors come
in linearly independent groups of four, the existence of a single one form $\d_i$ satisfying
(\ref{delti}) implies the existence of three more linearly independent
solutions, and thus (\ref{delti}) implies that either $\d_i=0$ or the base is
flat. Similarly, since $\theta$, $\theta^A$ are constant and
$\tilde{\n}J^A=0$, (\ref{deltii}) implies that
$[\tilde{\n}_i,\tilde{\n}_j]\theta_k=0$, and the existence of one
one form on the base satisfying this equation implies the existence of
three more linearly independent solutions. Hence either $\theta_j=0$
or the base is flat. If the base is not flat, then we must have
$\d_i=\theta_i=0$ and, given $m\neq0$, that $\theta=\theta^A=0$. Thus   
the solutions (\ref{su2sb}) preserve precisely four supersymmetries,
with a strictly $\su2s$ structure, when the base is a non-flat
hyperk\"{a}hler manifold.

\section{$SU(2)$ structure}
In this section we will study configurations in the theory admitting
an $SU(2)$ structure. We will first derive the constraints for the
existence of a particular choice of additional Killing spinor. We will
then derive neccessary and sufficient conditions for the existence of
eight Killing spinors fixed by the same $SU(2)$. Finally, we will
discuss some illustrative explicit solutions.

\subsection{One additional Killing spinor}
Consider an arbitrary linear combination of $(+,+)$ and $(-,+)$
spinors, which we denote by
\be
(\d+i\d^AT^A+\G^-(\theta+i\theta^AT^A))\e,
\end{equation}
and where at least one of the $\theta,\theta^A\neq0$. To derive the
constraints following from $\d\l=0$ for this spinor, we must impose
\be
\left[\Delta_{\l},\d+i\d^AT^A+\G^-(\theta+i\theta^AT^A)\right]\e=0.
\end{equation}
We have already calculated the
$\left[\Delta_{\l},\d+i\d^AT^A\right]\e$ terms, in the previous
section. Calculating the commutator
\be
\left[\Delta_{\l},i\theta^A\G^-T^A\right]\e,
\end{equation}
while entirely straightforward, is a tedious exercise. We will
therefore restrict attention to spinors with $\theta^A=0$, but
$\theta\neq0$. The $\theta$ dependent terms in the commutator, reduced
to canonical form, then become 
\bea
\left[\Delta_{\l},\theta\G^-\right]\e&=&\theta\Big[-5\pa_+\phi+2i
  J^{Aij}G_{+ij}T^A+2G_{+i5}\G^i+2G_{+-i}\G^-\G^i-5\pa_5\phi\G^-\nn &+&2iG_{5ij}J^{Aij}\G^-T^A\Big]\e.
\eea
We may immediately deduce
\be
G_{+i5}=G_{+-i}=\pa_5\phi=G_{5ij}^{(+)}=0.
\end{equation}
Then from the terms in $\left[\Delta_{\l},\d+i\d^AT^A\right]\e$, and
using $G^{(+)}_{5ij}=0$, we obtain
\bea
\pa_+\phi&=&0,\\\theta G_{+ij}J^{Aij}&=&-\frac{1}{4}(F^B_{ij}J^{Bij}\d^A-F^A_{ij}J^{Bij}\d^B).
\eea
Next, commuting the $-$ component of the Killing spinor equation, we
deduce
\bea
\pa_-\theta&=&0,\\
\pa_-\d&=&-\theta(2(5m+6he^{-4\phi})+\frac{1}{2}F^A_{ij}J^{Aij}),\\\pa_-\d^A&=&0.
\eea
From the $+$ component, we get
\bea
\pa_+\theta&=&0,\\
\pa_+\d&=&\theta\o_{+5+},\\\pa_+\d^A&=&0,\\\o_{++i}&=&0.
\eea
The 5 component gives
\bea
\pa_5\theta&=&0,\\\pa_5\d&=&\theta\o_{55+},\\\pa_5\d^A&=&0,\\\o_{5+i}&=&0,
\eea
and from the $i$ component, we get
\bea
\pa_i\theta&=&-\theta\pa_i\phi,\\\pa_i\d&=&0,\\\label{egg}\pa_i\d^A&=&-g\e^{ABC}A^B_i\d^C,\\\frac{1}{2}\e^{ABC}\Big(\frac{3}{5}F^{A(+)}_{ik}+F^{A(-)}_{ik}\Big)\d^BJ^{Ck}_{\;\;\;\;\;j}&=&\theta(\frac{1}{5}G^{(+)}_{+ij}+G^{(-)}_{+ij}-\o_{ij+}),\\\label{nof}\Big(\frac{1}{4}F^{A}_{kl}J^{Akl}+4he^{-4\phi}\Big)g_{ij}&=&F^{A(-)}_{ik}J^{Ak}_{\;\;\;\;\;j}.
\eea
As before, (\ref{egg}) implies that
$\e^{ABC}F^A_{ij}\d^C=0$. We may easily solve the differential
equations for $\theta$ to find
\be
\theta=ke^{-\phi},
\end{equation}
for some nonzero constant $k$; by a constant rescaling of the spinor, we may
take $k=1$. Furthermore, since
$A^{(+)}_{ik}B^{(-)k}_{\;\;\;\;\;j}$ is in general symmetric and
traceless on $i,j$, (\ref{nof}) implies that
\bea
F^A_{ij}J^{Aij}&=&-16he^{-4\phi},\\F^{A(-)}_{ij}&=&0.
\eea
\paragraph{Summary}Given the form (\ref{ymform}) of the Yang-Mills
fields, the following conditions, in addition to those of section 2,
are necessary and sufficient for a
bosonic configuration to admit the Killing spinors $\e$,
$(\d+i\d^AT^A+\theta\G^-)\e$, $\theta\neq0$. The functions $\theta$, $\d$ and $\d^A$
are required to satisfy 
\bea\label{ks}
\theta&=&e^{-\phi},\nn \pa_{-}\d&=&g,\nn\pa_i\d&=&0,\nn
\d^{A}&=&\d^A(x),\nn \pa_i\d^A&=&-g\e^{ABC}A^B_i\d^C,
\eea
The matter fields are constrained
according to
\bea\label{mf}
\phi&=&\phi(x),\nn
G_{(3)}&=&(4he^{-4\phi}-\frac{g}{2}e^{\phi})e^+\wedge e^-\wedge
e^5+e^+\wedge P_1^{(-)}\nn&+&e^-\wedge P^{(-)}_2+e^5\wedge
P^{(-)}_3+\frac{5}{2}\star_4d\phi,\nn
F^A&=&\frac{1}{2}F^{A(+)}_{ij}e^i\wedge e^j,\nn
F^A_{ij}J^{Aij}&=&-16he^{-4\phi},\nn \e^{ABC}F^B_{ij}J^{Cij}&=&0,\nn \e^{ABC}F^B_{ij}\d^C&=&0,
\eea
where $P_{1,2,3}^{(-)}$ are arbitrary anti selfdual two forms on the
base, and $\star_4$ denotes the Hodge dual on the base. Finally, there
are the following constraints on the independent components of the spin connection:
\bea\label{sc}
\o_{(\m\v)-}=\o_{-5i}&=&\o_{+5i}=\o_{++i}=0,\nn\o_{-+5}&=&-\frac{g}{2}e^{\phi},\nn\o_{-+i}&=&\o_{5i5}=\pa_i\phi,\nn\o_{+5+}&=&\theta^{-1}\pa_+\d,\nn \o_{55+}&=&\theta^{-1}\pa_5\d,\nn
\o_{+ij}&=&-\o_{ij+}=-P_1^{(-)},\nn \o_{-ij}&=&-P_2^{(-)},\nn
\o_{5ij}&=&-\o_{ij5}=-P_3^{(-)},\nn
\o_{ijk}J^{Ajk}&=&2gA^A_i-3\pa_j\phi J^{Aj}_{\;\;\;\;\;i},
\eea
and the remaining nonzero components may be read off from the
equalities in appendix C. It may be verified that it is sufficient to
impose the Bianchi identities for the forms and the field equation for
$G$, and all other equations of motion are implied by the existence of
the pair of Killing spinors. 

Clearly, demanding the existence of the
second Killing spinor significantly reduces the complexity of the
ansatz, and much more of the field content is fixed in terms of the
structure. Before considering some explicit examples, let us now turn
to the classification of all solutions admitting a strictly $SU(2)$ structure
with eight supersymmetries.

\subsection{$SU(2)$ structures with eight supersymmetries}
In this subsection, we will classify all configurations admitting
eight linearly independent Killing spinors sharing the common
projection $\G^{1234}\e=\e$ (note that this implies the common
projection $\G^{+-5}\e=-\e$). This is the maximal supersymmetry
compatible with a strictly $SU(2)$ structure. We will perform the
classification in full generality, and we thus relax the assumptions
made about the form of the Yang-Mills fields for this subsection. First
consider the vanishing of $\d\l$. Since this is linear in $\e$ and
contains no derivatives, it must vanish when acting on each of the
eight basis spinors individually. It is a simple matter to verify that
this implies that the matter fields are restricted to be of the
following form:
\bea
\phi&=&\phi(x),\nn F^{A}&=&\frac{1}{2}F^{A(-)}_{ij}e^i\wedge e^j, \nn
G&=&5(m+2he^{-4\phi})e^+\wedge e^-\wedge
e^5+\frac{1}{2}G_{-ij}^{(-)}e^-\wedge e^i\wedge
e^j+\frac{1}{2}G_{+ij}^{(-)}e^+\wedge e^i\wedge e^j\nn
&+&\frac{1}{2}G_{5ij}^{(-)}e^5\wedge e^i\wedge e^j +\frac{5}{2}\star_4
d\phi.
\eea
The vanishing of $F^A_{+\m}$, $F^A_{5\m}$ implies that locally we may choose a
gauge such that $A^A=A^A_ie^i$. Thus, acting on any of the eight
Killing spinors, the supercovariant derivative
reduces to the following form:
\bea
\md_-&=&\n_-+\frac{3}{2}(m+2he^{-4\phi})\G^{+5}-\frac{1}{2}\pa_i\phi\G^{+i}-m\G^+,\nn\md_+&=&\n_+-\frac{3}{2}(m+2he^{-4\phi})\G^{-5}-\frac{1}{2}\pa_i\phi\G^{-i}-m\G^-,\nn\md_5&=&\n_5+\frac{3}{2}(m+2he^{-4\phi})\G^{+-}+\frac{1}{2}\pa_i\phi\G^{5i}+m\G^5,\nn\md_i&=&\n_i+iF^A_{ij}\G^jT^A+\frac{1}{2}(G_{+ij}\G^{+j}+G_{-ij}\G^{-j}+G_{5ij}\G^{5j})\nn&+&(m+2he^{-4\phi})\G_i^{\;\;+-5}+\frac{1}{2}\pa_i\phi-igA^A_iT^A+m\G^i.
\eea
By assumption, there exist four Killing spinors of the form
$(\d+i\d^AT^A+\G^-(\theta+i\theta^AT^A))\e$, with four distinct
choices of the functions $\theta$, $\theta^A$, at least one of which
is nonzero in each case, and none of which is zero for all four. Let us evaluate  
$\left[\mathcal{D}_{\m},Q\right]\e$ for one of these Killing spinors. From the
$-$ component, we find the constraints
\bea
\pa_-\theta&=&\pa_-\theta^A=0,\nn
\pa_-\d&=&-2(5m+6he^{-4\phi})\theta,\nn\label{mimi}\pa_-\d^A&=&-2(5m+6he^{-4\phi})\theta^A.
\eea
Next, from the + component, we find
\bea
\pa_+\theta&=&\pa_+\theta^A=\o_{++i}=0,\nn \pa_+\d&=&-\theta\o_{++5},\nn\label{nmi}\pa_+\d^A&=&-\theta^A\o_{++5}.
\eea
The 5 component gives
\bea
\pa_5\theta&=&\pa_5\theta^A=\o_{5+i}=0,\nn\pa_5\d&=&-\theta\o_{5+5},\nn\label{mni}\pa_5\d^A&=&-\theta^A\o_{5+5}.
\eea
Finally, consider the $i$ component, and in particular, the
constraints derived from the vanishing of the $\G^{-i}\e$ term. These
read
\be\label{wohoo}
-\theta
F^A_{ik}J^{Ak}_{\;\;\;\;j}+\frac{1}{2}F^A_{ij}\theta^A+4h\theta
e^{-4\phi}g_{ij}+2he^{-4\phi}\theta^AJ^A_{ij}=0.
\end{equation}
Recall that each $F^A_{ij}$ is anti selfdual. Then extracting the
antisymmetric part of (\ref{wohoo}), and further decomposing in
selfdual and anti selfdual parts, we find
\bea
F^A_{ij}\theta^A&=&0,\nn hJ^A_{ij}\theta^A&=&0.
\eea
Since by assumption there exist three linearly independent Killing
spinors with different nonzero  $\theta^A$, these equations imply that $h=F^A=0$. Since
the Yang-Mills field strengths vanish, we may locally choose the gauge
so that also $A^A=0$. Then since each $T^A$ commutes with the
operators in the supersymmetry variations, we see that we may take four of the
Killing spinors to be $\e$, $iT^A\e$. It remains to determine the
other four
Killing spinors with nonzero $\theta$, $\theta^A$. Since $iT^A$
commute with the operators in the supersymmetry variations, once one
of the Killing spinors, $\e_K$, with nonzero $\theta$, $\theta^A$ is
determined, the other three may be taken to be $iT^A\e_K$. We thus
only need to determine the single Killing spinor $\e_K$. The remaining constraints from
the $i$ component of the Killing spinor equation are
\bea
\pa_i\d&=&\pa_i\d^A=0,\nn
\pa_i\theta&=&-\pa_i\phi\theta,\nn\pa_i\theta^A&=&-\pa_i\phi\theta^A,
\nn G_{ij+}&=&\o_{ij+}.
\eea
We may easily solve for the differential equations for $\theta$, $\theta^A$ to find
\bea
\theta&=&ke^{-\phi},\nn\theta^A&=&k^Ae^{-\phi},
\eea
for some constants $k$, $k^A$, at least one of which is nonzero. Now,
note from the differential equations for $\d$, $\d^A$ that if $k=0$,
then $\d$ is constant, and if $k^A=0$, then $\d^A$ is constant. These
constants may be taken to be zero, since we are free to add to $\e_K$
a constant multiple of the four Killing spinors $\e$, $iT^A\e$. Let us
thus write
\bea
\d&=&k\d^{\prime},\nn\d^A&=&k^A\d^{A\prime},
\eea
with no sum on $A$ in the second of these equations. Now for each
nonzero $k$, $k^A$, the associated $\d^{\prime}$, $\delta^{A\prime}$
obey the same set of differential equations, which schematically are
\be
\pa_{\m}f=B_{\m}.
\end{equation}
Locally the solution exists and is unique, up to a constant which as
before may be taken to be zero. Hence
\bea
\d&=&kf(u,v,z,x),\nn\d^A&=&k^Af(u,v,z,x).
\eea
We may easily fix the $v$ dependence of $f$ from equation
(\ref{mimi}), finding
\be
f(u,v,z,x)=gv+\hat{f}(u,z,x).
\end{equation}
We must impose $\pa_if=0$, but the function $\hat{f}$ is otherwise
arbitrary. Making a choice for it determines $\o_{++5}$ and $\o_{5+5}$
through equations (\ref{nmi}) and (\ref{mni}). Thus 
\be
\e_K=(kgv+k\hat{f}+iT^Ak^A\hat{f}+e^{-\phi}\G^-(k+ik^AT^A))\e,
\end{equation}
and the remaining three Killing spinors are $iT^A\e_K$. By taking
linear combinations with constant coefficients of these four spinors,
we may in fact choose the set of four Killing spinors to be
\be
\e^{\prime}_K=(gv+\hat{f}(u,z,x)+e^{-\phi}\G^-)\e,
\end{equation}
and $iT^A\e^{\prime}_K$. We have thus determined all eight Killing
spinors, and the necessary and sufficient conditions for their existence. It is intriguing to note that all eight may be generated by
repeated application of the three matrices $iT^1$, $iT^2$ and
$(gv+f^{\prime}(u,z,x)+e^{-\phi}\G^-)$ to $\e$, and also that the
third one of these matrices is precisely of the form we derived in the previous
subsection, for a single additional Killing spinor. Perhaps for
all configurations with an $SU(2)$ structure, the additional Killing
spinors are of this form; that is, perhaps these are the only three
matrices defining an $SU(2)$ structure which, together with their
products, can commute with $\Delta_{\l}$ and $\md_{\mu}$, when acting on
$\e$. We have not verified this conjecture. However, if it is true
that there are only a few matrices $Q$ that can generate additional
Killing spinors $Q\e$, then the problem of performing a fully refined
classification will not be as formidable as it might appear. It will
be interesting to explore this point in the future, and in other
contexts. We will now turn to some explicit examples.

\subsection{Examples}
To illustrate the case of an $SU(2)$ structure, let us first see how
known $AdS_3\times\mathcal{M}_4$ solutions of the theory arise in our
formalism. This will also illustrate the effect of additional Killing spinors on the
intrinsic torsion of the structure. We will then generalise the
construction, to obtain new $AdS_3$ solutions.  
\subsubsection{$AdS_3\times\mathcal{M}_4$ solutions}
Let us make the following ansatz for a second independent Killing
spinor:
\be
\e_K=(gv+\G^-)\e.
\end{equation}
This choice solves all the differential constraints on $\theta$,
$\delta$; also, it implies that $\o_{+5+}=\o_{55+}=0$, and $\phi$=constant; without loss
of generality, when $\phi=$constant, by constant rescalings of the forms and the couplings
we may take $\phi=0$. We will take
the metric to be of the direct product form
\be
ds^2=L^{-1}(z)dudv+C^2(z)dz^2+h_{MN}(x)dx^Mdx^N.
\end{equation}
For this choice of metric, the constraint $\o_{-+5}=-g/2$ is equivalent to
\be
\pa_zL=-gLC,
\end{equation}
from Appendix C, which is solved by
\be
L=(gz)^{-2},\;\;C=2L^{1/2},
\end{equation}
and so by rescaling $z$ the metric in the $+-5$ directions may be written as 
\be
\frac{4}{g^2}(z^2(-dt^2+dy^2)+\frac{dz^2}{z^2}),
\end{equation}
which is the metric on $AdS_3$ with $AdS$ length $2g^{-1}$. Our choice
of metric implies that $P^{(-)}_{1,2,3}=0$, so the fluxes are
\bea
G_{(3)}&=&(4h-\frac{g}{2})e^+\wedge e^-\wedge e^5,
\nn F^A&=&f^{AB}J^B,
\eea
where
\bea
f^{[AB]}&=&0,\nn f^{AA}&=&-4h.
\eea
Since $G$ is closed and coclosed, the Bianchi identity for $G_{(4)}$ is
automatically satisfied, while the field equation for $G_{(4)}$ is
\be\label{geqn}
-8h(4h-\frac{g}{2}) -f^{AB}f^{AB}=0.
\end{equation}
We may rewrite the final constraint of (\ref{sc}) in covariant form as
\be
\n_iJ^A_{jk}+g\e^{ABC}A^B_iJ^C_{jk}=0,
\end{equation}
or equivalently,
\be
A^A_i=-\frac{1}{8g}\e^{ABC}J^{Bjk}\n_iJ^C_{jk}.
\end{equation}
Imposing the Yang-Mills Bianchi identity then implies that 
\bea
\label{ein}f^{AB}J^B_{ij}&=&\frac{1}{2g}J^{Akl}R_{klij}\equiv\frac{1}{g}\mr^A_{ij}.
\eea
Taking the of commutator of two $SU(2)$ covariant derivatives acting
on $J^A$ we may obtain
\be
J^{Ak}_iR_{kj}=\mr^A_{ij}+\e^{ABC}J^{Bk}_i\mr^C_{kj},
\end{equation}
where $R_{ij}$ is the Ricci tensor, and hence that
\bea
R&=&4gf^{AA}=-16hg, \\R_{ij}&=&\frac{R}{4}g_{ij},
\eea
where $R$ is the scalar curvature of the base. Hence the Yang-Mills
Bianchi identity implies that we must choose the base to be
Einstein (with negative scalar curvature when $hg>0$). The Yang-Mills fields are then determined by
(\ref{ein}), and once $f^{AB}$ is determined, (\ref{geqn}) fixes $h$
in terms of $g$. There are no further constraints, and all remaining field
equations are identically satisfied.

\paragraph{Two Killing spinors} 
Let us take the base to be $H^4$, equipped with the metric
\be
ds_4^2=\frac{4\m^2}{(1-r^2)^2}(dr^2+\frac{r^2}{4}\d_{AB}\s^A\s^B),
\end{equation}
where the $\s^A$ are the right invariant one forms on an $S^3$,
$d\s^A=-\frac{1}{2}\e^{ABC}\s^B\s^C$. Let us choose the vierbeins to be
\be
e^1=\frac{\m r}{1-r^2}\s^1,\;\;e^2=\frac{\m
  r}{1-r^2}\s^3,\;\;e^3=\frac{\m
  r}{1-r^2}\s^2,\;\;e^4=-\frac{2\m}{1-r^2}dr.
\end{equation}
Then we may simply read off the full solution; we find that
$g=\frac{28}{3}h$, $\m^2=\frac{7}{g^2}$, and the bosonic fields are given by
\bea
ds^2&=&\frac{1}{g^2}ds^2(AdS_3)+\m^2ds^2(H^4),\nn
\phi&=&0,\nn
G_{(4)}&=&\frac{1}{2g\m^2}e^{1234},\nn
F^A&=&-\frac{1}{g\m^2}(e^{A}\wedge e^4+\frac{1}{2}\e^{ABC}e^{B}\wedge
e^C), \nn A^A&=&\frac{r^2}{g(r^2-1)}\s^A.
\eea
This solution admits precisely the two Killing spinors $\e$ and
$\e_K=(gv+\G^-)\e$, and it may readily be verified that both
of these spinors are null, as are their sum and difference. Uplifted
to eleven dimensions, this
solution is the $AdS$ fixed point of the near-horizon limit of an
$M5$ brane wrapped on a Cayley 4-cycle in a $Spin(7)$ manifold \cite{wrap}.

\paragraph{Four Killing spinors}
Now suppose that in addition to the Killing spinors $\e$, $\e_K$, we
demand the existence of the Killing spinor $T^1\e$. This imposes
$F^2=F^3=0$, also it implies the
existence of the fourth Killing spinor
$\e_{K}^{\prime}=T^1(gv+\G^-)\e$. Since $A^2=A^3=0$, we now have
\be
\tilde{\n} J^1=0,
\end{equation}
so the base must be chosen to be Einstein-K\"{a}hler, with
$F^1=g^{-1}\mr$, where 
$\mr$ the Ricci form of the base. From the four form field equation we
find $g=12h$. The four spinors $\e$, $T^1\e$, $\e_{K}$, $\e_K^{\prime}$ are
null. However we know from \cite{us} that these solutions admit
timelike spinors. It is easily checked that the four Killing spinors
$\e\pm\e_{K}^{\prime}$, $T^1\e\pm\e_K$ are timelike. Uplifted to
eleven dimensions, the metric is that of the $AdS$ fixed point of the
near-horizon limit of an $M5$ wrapped on a K\"{a}hler 4-cycle in a
Calabi-Yau 4-fold \cite{wrap}. 

\paragraph{Eight Killing spinors}
Finally, suppose that in addition to the four Killing spinors of the
previous paragraph, we demand the existence of the Killing spinor
$T^2\e$. This imposes  
\bea
F^A&=&h=0,\\G_{+-5}&=&\o_{-+5}=-\frac{g}{2},\\
\n J^A&=&0,
\eea
and implies the existence of the additional Killing spinors
$T^3\e$, $T^2(gv+\G^-)\e$, $T^3(gv+\G^-)\e$.
The three $J^A$ are required to be covariantly constant on the base,
which must thus be taken to be
Calabi-Yau. As in the previous example, these solutions admit both
timelike and null Killing spinors. Lifted to ten dimensions, they are
just the familiar $AdS_3\times S^3\times\mathcal{M}_4$ solutions of
ten dimensional supergravity.

The $AdS$ examples we have given clearly illustrate the effect that
the additional independent Killing spinors have on the torsion of the
G-structure, which is $SU(2)$ in each case. The base is progressively
restricted from Einstein to Einstein-K\"{a}hler to Calabi-Yau, together
with an associated truncation of the Yang-Mills fields from $SU(2)$ to
$U(1)$ to zero, as additional Killing spinors are incorporated.  

\subsubsection{Generalisations: membranes with $AdS_3$ worldvolume}
Let us now consider generalising the known solutions given above, by
allowing for a non-constant dilaton. In fact, the only additional truncation of
the matter fields we will make, beyond that implied by the required
supersymmetry, is to set $P^{(-)}_{1,2,3}=0$. Let us  demand the
existence of the second Killing spinor 
\be
\e_K=(gv+e^{-\phi}\G^-)\e.
\end{equation}
This implies that $\o_{+5+}=\o_{55+}=0$. Let us make the
  metric ansatz 
\be
ds^2=L^{-1}(x,z)dudv+C^2(x,z)dz^2+h_{mn}(x)dx^Mdx^N.
\end{equation}
Now let us solve the
constraints 
\be
\o_{-+i}=\o_{5i5}=\pa_i\phi.
\end{equation}
These read
\be
\frac{1}{2L}\pa_iL=-\frac{1}{C}\pa_iC=\pa_i\phi,
\end{equation}
which are solved by
\bea
L&=&e^{2\phi}\tilde{L}(z),\nn C&=&e^{-\phi}\tilde{C}(z).
\eea
Next, the constraint
\be
\o_{-+5}=-\frac{g}{2}e^{\phi},
\end{equation}
 becomes
\be
\pa_z\tilde{L}=-g\tilde{C}\tilde{L},
\end{equation}
which as before is solved by
\be
\tilde{L}=(gz)^{-2},\;\;\;\tilde{C}=2\tilde{L}^{1/2}.
\end{equation}
Given our metric ansatz, all the remaining constraints
on the spin connection are satisfied, except for
\be
\o_{ijk}J^{Aij}=2gA^A_i-3\pa_j\phi J^{Aj}_{\;\;\;\;\;i}.
\end{equation}
Conformally rescaling the base according to
\be
h_{(4)}=e^{3\phi}\tilde{h}_{(4)},
\end{equation}
we find
\be
\tilde{\o}_{ijk}\tilde{J}^{Ajk}=2gA^A_i,
\end{equation}
where $J^A_{ij}=e^{3\phi}\tilde{J}^A_{ij}$, $\tilde{\o}_{ijk}$ is the
spin connection on the base with metric $\tilde{h}_{(4)}$, and for the
remainder of this discussion all indices on ``tilded'' objects are raised with
$\tilde{h}^{ij}$. The seven metric is thus
\be
ds^2=4(ge^{\phi})^{-2}\Big(z^2(-dt^2+dy^2)+\frac{dz^2}{z^2}\Big)+e^{3\phi}d\tilde{s}^2,
\end{equation}
the warped product of $AdS_3$ with some four-manifold. Next, imposing the Bianchi identity for $F^A$ (which we recall in our
conventions for the field strength is of the modified form
$D(e^{\phi}F^A)=0$), we find that
\be\label{bbnn}
F^A_{ij}=(ge^{\phi})^{-1}\tilde{\mr}^A_{ij},
\end{equation}
and as before that
\be
\tilde{R}=-16hg,
\end{equation}
where the conformally rescaled metric $\tilde{h}$ is either Einstein,
K\"{a}hler-Einstein or Calabi-Yau, depending on whether the Yang-Mills
fields are $SU(2)$, $U(1)$ or zero. It remains to impose the Bianchi
identities and field equations for $G$. It is trivially verified that
the four-form Bianchi identity, $d(e^{-2\phi}G_{(4)})=0$, is
satisfied. Finally, the field equation for $G_{(4)}$ reduces to a
single equation for $\phi$:
\be\label{curvybase}
\frac{5}{2}(\tilde{\n}^2\phi+5\tilde{h}^{ij}\pa_i\phi\pa_j\phi)-4hg+h^2e^{-5\phi}
(32+\frac{16}{n})=0,
\end{equation}
where $n=3$ for $SU(2)$ Yang-Mills fields and $n=1$ for $U(1)$. When
the Yang-Mills fields vanish, we must take $h=0$. The 
examples in the previous subsection are clearly $\phi=0$ solutions of this equation. When
$\phi$ is not constant, by making the
definition
\be
\phi=\frac{1}{5}\log\Big(f+\frac{4h}{g}\Big(2+\frac{1}{n}\Big)\Big),
\end{equation}
we find that $f$ obeys
\be
\n^2f-8hgf=0.
\end{equation}
As a first example of a more general solution of this form, take the
conformally rescaled base to be $H^4$, with squared radius
$\frac{3}{4hg}$, and metric 
\be
d\tilde{s}^2=\frac{3}{4hg}(dr^2+\sinh^2dd\O_3^2).
\end{equation}
We take $f=f(r)$, and thus obtain
\be\label{FFFF}
f^{\prime\prime}+3\coth rf^{\prime}-6f=0.
\end{equation}
Making the change of variable $u=\tanh^2r$, and defining
$f=(1-u)^{\a}\psi$, $\a=(3+\sqrt{33})/4$, converts this to
hypergeometric form, 
\be
u(1-u)\frac{d^2\psi}{du^2}+[c-(1+a+b)u]\frac{d\psi}{du}-ab\psi=0,
\end{equation}
with $a=(3+\sqrt{33})/4$, $b=(5+\sqrt{33})/4$, $c=2$. Depending on the
solution we choose for $\psi$, we find that the metric is  singular either at
$r=0$ or $r=\infty$, or both.    
The $SU(2)$ Yang-Mills fields may be read off from
(\ref{bbnn}). Imposing regularity at infinity, we choose
$\psi=F(a,b;1+a+b-c;1-\tanh^2r)$; we find that for large $r$, the
solution asymptotes to our previous $AdS_3\times H^4$ example. To
investigate the behaviour near $r=0$, we define a new radial
coordinate $\rho=r^{2/5}$; the metric approaches that of a cone over
$AdS_3\times S^3$,
\be
ds^2=d\rho^2+\rho^2(R_1^2ds^2(AdS_3)+R^2_2ds^2(S^3)),
\end{equation}
and the dilaton blows up. The full solution preserves two supersymmetries. 

As a second example of a more general solution of this form, take the
conformally rescaled base metric to be flat,
\be
d\tilde{s}^2=dr^2+r^2d\O_3^2.
\end{equation}
This choice implies that $F^A=h=0$. Let us set $f=f(r)$, so that up to an irrelevant multiplicative constant,
\be
f=1+\frac{a}{r^2}.
\end{equation}
The seven dimensional metric is thus
\be
ds^2=4g^{-2}\Big(1+\frac{a}{r^2}\Big)^{-2/5}ds^2(AdS_3)+\Big(1+\frac{a}{r^2}\Big)^{3/5}ds^2(\mbb^4).
\end{equation}
Clearly, for large $r$ the metric becomes that of
$AdS_3\times\mbb^4$. To investigate the behaviour near $r=0$, let us
define $u=\frac{5}{2}a^{1/2}r^{2/5}$. Near $r=0$, the metric becomes
\be
ds^2=du^2+u^2\Big(\frac{16}{25g^2}ds^2(AdS_3)+\frac{4}{25}ds^2(S^3)\Big),
\end{equation}
which is again that of a cone over $AdS_3\times
S^3$. Again, the dilaton blows up as we approach $r=0$. The full
solution preserves eight supersymmetries.

\section{$\mathbb{R}^5$ structure}
Next consider an arbitrary linear combination of $(+,+)$ and $(+,-)$
spinors, which we denote by
\be
(\d+i\d^AT^A+\d_i\G^i)\e,
\end{equation}
and where at least one of the $\d_i\neq0$. It is convenient to
introduce an orthonormal basis on the four dimensional base. Defining
$H^2=\d_i\d^i$, we choose the basis
\be
e^A_i=H^{-1}J^A_{ji}\d^j,A=1,..,3,\;\;\;e^4_i=H^{-1}\d_i.
\end{equation}
In this basis, we have $\e^{1234}=1$. We will thus write
$\e^{ABC4}=\e^{ABC}$. To avoid potential confusion with this mixing of
spacetime and Yang-Mills indices, we will always
place Yang-Mills indices on the $J^A$ and the $F^A$ ``up'' and
spacetime indices ``down''. However we will make no distinction
between ``up'' and ``down'' indices on $\e^{ABC}$. The
components of the $J^A$ in this basis are
\be
J^A_{B4}=-\d^{AB},\;\;\;J^A_{BC}=-\e^{ABC}.
\end{equation}
For computational convenience, we will restrict attention to Killing
spinors which are not of the most general form compatible with an
$\mathbb{R}^5$ structure, but which are rather of the form
$\d_i\G^i\e=H\G^4\e$. 
Evaluating
\be
\left[\Delta_{\l},\d_i\G^i\right]\e=0,
\end{equation}
and employing the constraints of section three, we find the conditions
\bea
G_{-ij}&=&G_{ijk}=0,\nn
\e^{ABC}F^A_{BC}&=&20(m+2he^{-4\phi}),\nn\e^{ABC}F^B_{4C}&=&\e^{ABC}G_{5BC}.
\eea
Next, using $\o_{\m\;\;\;\;\v}^{\;\;\;4}=-\n_{\m}e^4_{\v}$, we
deduce from commuting the $-$ component of the Killing spinor equation that
\be
\pa_-H=0.
\end{equation}
From the $+$ component we find
\bea
\pa_+H&=&h=0,\nn\o_{+4A}&=&G_{+4A},\nn
G_{45A}&=&\frac{1}{2}F^{B}_{BA}.
\eea
The 5 component yields
\bea
\pa_5H&=&0,\nn
\o_{54A}&=&G_{54A},
\eea
while the $B$ component gives
\bea
\pa_BH=F^A_{BC}&=&g=0,\nn
\o_{B4A}&=&\pa_4\phi g_{AB},\nn
\e^{ACD}\o_{BCD}&=&\frac{3}{2}\pa_4\phi g_{AB}+\e^{ABC}\pa_C\phi.
\eea
Since the gauge coupling $g$ is required to be zero, an $\mathbb{R}^5$
structure of the form we have assumed is only admitted in the ungauged
theory. The remaining constraints may be derived from the 4 component
of the Killing spinor equation, and we find them to be of the form
\bea
H&=&\mbox{const},\nn
\pa_5\phi&=&0,\nn\e^{ABC}F^B_{4C}&=&G_{5ij}=0,\nn\o_{4A4}&=&\pa_A\phi.
\eea
We have thus derived the complete set of additional constraints on a
bosonic configuration for it to admit the second Killing spinor
$\d_i\G^i\e$. However, there are no solutions of this form in the
gauged theory, since the coupling is zero. We have also verified that there are no vacuum
solutions, or solutions with eight Killing spinors, with an $\mbb^5$
structure in the gauged theory. We will not consider this case any further. 

\section{Identity Structure}
Making a more general ansatz for a second Killing spinor than that
consistent with an $(SU(2)\ltimes\mathbb{R}^5)\times\mathbb{R}$,
$SU(2)$ or $\mathbb{R}^5$ structure implies that the structure group
is reduced to the identity. The generic such Killing spinor is
parameterised by sixteen real functions. Computing the constraints
associated with the existence of such a spinor is a lengthy 
computational exercise. To illustrate the case of an identity structure, we will instead restrict attention to Killing spinors of
the simpler form
\be\label{spinspin}
\theta_i\G^{-i}\e,
\end{equation}
that is, pure $(-,-)$ spinors, in the language of the introduction. 

\subsection{Constraints}
As
in the previous section, we introduce an orthonormal basis on the
base; as before, we define $H^2=\theta_i\theta^i$, and choose
\be
e^A_i=H^{-1}J^A_{ji}\theta^j,\;\;\;e^4_i=H^{-1}\theta_i.
\end{equation}
Commuting the dilatino variation, we may derive
\bea
\pa_+\phi=\pa_i\phi&=&G_{+5i}=G^{(-)}_{+ij}=0,\nn
G_{+-5}&=&\frac{1}{2}F^A_{4A},\nn
G_{45A}&=&\frac{1}{2}\e^{ABC}F^B_{4C}.
\eea
Now, from the $+$ component of the Killing spinor equation, we get
\be
\pa_+H=\o_{++\m}=F^A_{4A}=\e^{ABC}F^B_{4C}=\o_{+ij}^{(-)}=0.
\end{equation}
The 5 component gives
\bea
\o_{5+5}=\o_{5+i}&=&\o_{554}=\o_{54A}=0,\nn \pa_5(\log
H+\phi)&=&4he^{-4\phi}.
\eea
Next from the $-$ component we find
\bea
\pa_-H=\o_{-+i}=G_{-+i}&=&G_{ijk}=G_{-ij}=0,\nn \e^{ABC}G_{5BC}&=&F^B_{BA}.
\eea
The $A$ component yields the constraints
\bea
\pa_AH=\o_{A54}&=&\o_{AB4}=F^A_{4B}=0, \nn \o_{ij+}&=&G_{ij+}.
\eea
Finally the 4 component gives
\be
\pa_4H=\o_{44A}=0.
\end{equation}
These, in addition to the constraints of section 2, are the full
set of neccessary and sufficient conditions on the bosonic fields of
the theory for the existence of a second Killing spinor of the form
(\ref{spinspin}). We see that demanding the existence of such a
Killing spinor, with its associated identity structure, implies a
radical simplification of the general problem. In fact, we will make
one further simplifying assumption. By inspection of the constraints
on the spin connection,
we see that $de^5=0$ if and only if $\e^{ABC}G_{5BC}=F^B_{BA}=0$; we
will assume that $G_{5AB}=0$. Then since $de^5=0$, we may always
choose our local coordinates such that $e^5=dz$, that is, $C=1$,
$\s=0$. 

\paragraph{Summary}
Given the assumption that $G_{5AB}=0$, demanding the existence of the
second Killing spinor (\ref{spinspin}) implies that we may choose coordinates
such that $e^5=dz$, and that the matter fields and the
function $H$ are
of the following form:
\bea 
\phi&=&\phi(z), \nn H&=&H(z), \nn G_{(3)}&=&e^+\wedge Q^{(+)}, \nn
F^A&=&\frac{1}{2}F^A_{BC}e^B\wedge e^C,
\eea
where $Q^{(+)}$ is a selfdual form on the base. There are the
following constraints on $\phi$, $H$ and $F^A$:
\bea
\pa_z\log(He^{\phi})&=&4he^{-4\phi},\nn F^A_{AB}&=&0, \nn
\e^{ABC}F^A_{BC}&=&10(\pa_z\log H+2m).
\eea
The only nonzero independent spin connection components are the
following:
\bea
\o_{-+5}&=&-\pa_z\log
H,\nn\o_{+ij}&=&\o_{+ij}^{(+)},\nn\o_{ij+}&=&Q_{ij},\nn
\o_{454}&=&-\pa_z\log H,\nn \o_{A5B}&=&(\frac{3}{2}\pa_z\log
H+5m)g_{AB}+\frac{1}{2}\e^{CBD}F^C_{AD}-\frac{1}{4}\e^{ADE}F^B_{DE},\nn
\o_{iAB}&=&-g\e^{ABC}A^C_i.
\eea
The degree of simplification in this case is quite remarkable.

\subsection{Examples} 
Note that since
$\o_{(\m\v)+}=0$, $\frac{\pa}{\pa u}$ is Killing. We will for
simplicity seek solutions with $\o_{ij+}=G_{ij+}=0$. Then note that performing the rescalings
\be
e^+=H\tilde{e}^+,\;\;\;e^-=H\tilde{e}^-,\;\;\;e^5=H\tilde{e}^5,\;\;\;e^4=H\tilde{e}^4,
\end{equation}
the constraints on $\o_{+-5}$ and $\o_{454}$ imply that
$d\tilde{e}^-=d\tilde{e}^+=d\tilde{e}^4=d\tilde{e}^5=0$; thus locally
we may introduce coordinates such that 
\bea
\tilde{e}^+&=&du,\nn\tilde{e}^-&=&dv,\nn\tilde{e}^4&=&dr,\nn\tilde{e}^5&=&dz.
\eea
Note that we have redefined the $z$ coordinate.  Furthermore, since
$u$ is Killing, the base is independent of $u$. We have thus solved all the constraints on the spin connection except
\bea \label{ppaa}
\o_{A5B}&=&(\frac{3}{2}H^{-1}\pa_z\log
H+5m)g_{AB}+\frac{1}{2}\e^{CBD}F^C_{AD}-\frac{1}{4}\e^{ADE}F^B_{DE},\\\label{ppaap}
\o_{iAB}&=&-g\e^{ABC}A^C_i,
\eea
which we have written in terms of our new $z$ coordinate. 

The Yang-Mills Bianchi identities together with (\ref{ppaap}) then
imply that, as in the case of the $SU(2)$ structure,
\be
e^{\phi}F^A_{ij}=\frac{1}{g}\mr^A_{ij},
\end{equation}
where the $\mr^A_{ij}$ are the curvatures of the right hand spin bundle
of the base with metric
\be
ds^4=H^2(z)dr^2+\d_{AB}e^Ae^B.
\end{equation}
Recall that $F^A_{4B}=0$, and in addition to (\ref{ppaa}),
(\ref{ppaap}) we have the additional constraints, written in terms of
our new $z$ coordinate: 
\bea
\label{near}\pa_z\log(He^{\phi})&=&4hHe^{-4\phi},\\ F^A_{AB}&=&0, \\
\label{nearly}\e^{ABC}F^A_{BC}&=&10(H^{-1}\pa_z\log H+2m).
\eea
Let us now consider some explicit examples.

\paragraph{Two Killing spinors}
Recall that since the scalar curvature of the base is given by
$R=J^{Aij}\mr^A_{ij}=gJ^{Aij}e^{\phi}F^A_{ij}=-ge^{\phi}\e^{ABC}F^A_{BC}$.
Equation (\ref{nearly}) implies that $R$ is a function only of $z$. Let us look for a solution with a base of the form
\be
H^2(dr^2+\rho^2(z)ds^2(S^3)),
\end{equation}
so that 
\be\label{belderrig}
F^A_{BC}=-(H\rho)^{-2}e^{-\phi}\e^{ABC}=\frac{5}{3}(H^{-1}\pa_z\log
H+2m)\e^{ABC}.
\end{equation}
We will seek a solution with $h=0$, so that (\ref{near}) implies that
up to an irrelevant multiplicative constant, $H=e^{-\phi}$. Then
writing $H=f(z)e^{gz/5}$, the second equality of
(\ref{belderrig}) becomes 
\be
-\rho^{-2}=\frac{5}{3}\pa_z\log f
\end{equation}
Next, using the expression for the spin connection given in Appendix
C, (\ref{ppaa}) is
\be
\pa_z(3\log\rho+10\log f)=0,
\end{equation}
and hence
\be
\rho^3=\b f^{-10},
\end{equation}
for some constant $\b$. Hence we find that
\be
f^{-20/3}=\frac{4}{\b}z+\a,
\end{equation}
and so shifting $z$ to eliminate the constant $\a$, the metric and
dilaton may be written as
\bea
ds^2&=&\Big(\frac{4z}{\b}\Big)^{-3/10}e^{2gz/5}(-dt^2+d\mbx^2+dz^2+4zds^2(S^3)),\nn
e^{\phi}&=& \Big(\frac{4z}{\b}\Big)^{3/20}e^{-gz/5}. 
\eea
This describes the near-horizon limit of a IIB fivebrane wrapped on an
associative three-sphere \cite{ach}.

\paragraph{Four Killing spinors} Let now us seek $U(1)$ solutions; this is
equivalent to demanding the existence of the Killing spinor
$iT^1\e$. Then we find that the $U(1)$ gauge field obeys $F^1_{1A}=0$,
and hence that $\o_{A51}=-H^{-1}\pa_z\log H\d_{1A}$, $\o_{1A5}=\o_{iA1}=0$. Hence the configuration
also admits the Killing spinor $H\G^-\G^1$. We rescale $e^{1}$, so that
locally we can write the metric as
\be
ds^2_7=H^2ds^2(\mbb^5)+ds^2(\mathcal{M}_2).
\end{equation}
Consider solutions with $\phi=0$. Then $H=-(4hz)^{-1}$,
$F_{23}=16h-g$, and hence the scalar curvature of $\mathcal{M}_2$ is
\be
R=-\frac{g}{2}(16h-g),
\end{equation}
and so $\mathcal{M}_2$ is either $\mbb^2$, $S^2$ or $H^2$, and
is independent of the coordinates on the five othogonal
directions. Hence we must impose $\o_{A5B}=0$, $A$, $B=2,3$, and
therefore $12h=g$. Thus the solution is the $AdS_5\times H^2$ solution
of Maldacena and Nunez \cite{carlos}, for which we have displayed the identity
structure.

\paragraph{Eight Killing spinors}
Finally, let us look for solutions for which the Yang-Mills
fields vanish. This is equivalent to demanding the existence of the
eight Killing spinors $\e$, $iT^A\e$, $H\G^-\G^i\e$. The vanishing of
the Yang-Mills fields implies that $\o_{A5B}=-\pa_z\log
H\d_{AB}$. Then rescaling the $e^A$ by $H$, we see that locally the
metric can be put in the form
\be
ds^2=H^2\eta_{\m\v}dx^{\m}dx^{\v}.
\end{equation}
Furthermore we must require
\bea
\pa_z\log H&=&-2mH,\nn
\pa_z\log(He^{\phi})&=&4Hhe^{-4\phi}.
\eea
Consider first the case of vanishing topological mass,
$h=0$. Then up to irrelevant constants,
$H=e^{-\phi}=\exp{\frac{gz}{5}}$. This solution is nothing but the
reduction to seven dimensions of the linear dilaton solution in ten.

When $h\neq0$, let us look for a solution with $\phi=0$. We thus must
have $16h=g$, $H^2=(4hz)^{-2}$. This is just the maximally
supersymmetric $AdS_7$ solution of the theory.

\section{Conclusions and outlook}
In this work, a systematic formalism for performing complete
G-structure classifications of supersymmetric bosonic configurations in
supergravity theories has been presented, and illustrated in the
context of $d=7$, $SU(2)$ gauged supergravity. The key notion for
organising the classification is that of the common isotropy group of
the additional Killing spinors. The formalism has been used to
derive a set of constraints associated wth the existence of various
additional Killing spinors, and these constraints have been exploited
to derive numerous explicit solutions. The emphasis in this paper has
been on illustrating the formalism, and no great effort has been made
to derive new solutions; clearly, there is scope for a more careful
analysis of the constraints in the future.

An important technical point that has been exploited throughout is
that in performing G-structure classifications, it is unneccessary, and
inefficient, to use spinor bilinears to derive the constraints;
for more computational efficiency, it is much better to work
directly with a specific spinor defined by a particular set of
projections. The second key point that has been used is that given a
single Killing spinor, any other spinor may be constructed from it by acting
with a matrix $Q$ in the Clifford algebra. The problem of
completely classifying all supersymmetric configurations in a given
supergravity may then be restated as determining all possible sets of
such matrices $Q$ which commute with the operators in the fermion
supersymmetry transformations, when acting on the fiducial Killing
spinor. 

The chief advantage of the G-structure formalism used here over other
classification techniques is that the constraints one derives for the
existence of the desired supersymmetries take the form of explicit
algebraic constraints on the spin connection and the matter fields,
supplemented by first order differential conditions on the functions
parameterising any additional Killing spinors. Presenting the
neccessary and sufficient conditions for the existence of $N$ Killing
spinors in this form naturally lends itself to exploiting the
classification in the construction of explicit solutions, rendering 
G-structure classifications very useful for practical
applications. 

The only drawback of the formalism is, of course, the ammount of
computational effort required to compute the constraints in a theory
of the complexity of the one studied here. It is to be expected that
simpler theories, such as those with eight supercharges, would be much
more tractable from the point of view of perfoming a refined
classification in full generality, using the techniques of this paper;
this will be interesting to
investigate. And it is of course possible that there is a yet
more efficient way of computing the constraints. The approach employed
here is ``bottom up'', in the sense that one first assumes the
existence of a single Killing spinor, and then incorporates additional
Killing spinors iteratively. For the analysis of more complicated
supergravities, it would be very useful to have a complementary ``top
down'' approach, so that one could start with maximal supersymmetry
and weaken the constraints progressively, in a controlled
fashion. This would be particularly useful for classifying
configurations admitting more than one half supersymmetry, since in
this case the structure group is neccessarily the identity, and this
is the most complicated case to analyse using the iterative bottom up
approach.  

An obvious application of the formalism presented here is to the
long-standing problem of completely classifying supersymmetric
configurations in $d=11$. The possible structure groups in
$d=11$ will coincide with the possible holonomy groups; these have
been classified in \cite{jose}, and there are eighteen distinct
possibilites. A G-structures analysis of the maximal structure groups
($SU(5)$ and $(Spin(7)\ltimes\mbb^8)\times\mbb$) has been given in
\cite{gaunt1}, \cite{gaunt3}. The first step in performing the
complete classification will be to construct, from a fiducial Killing
spinor, the spaces of spinors fixed by each of the other sixteen groups, as was
done in the context of this paper in the introduction. Then it should
certainly be practical to completely analyse the thirteen additional
structure groups fixing at most eight Killing spinors; analysing the two groups
($SU(2)$ and $\mbb^9$) fixing at most sixteen will require
considerably more
effort. Finally, completely classifying the most generic case of an
identity structure (which fixes at most thirty-two Killing spinors)
will, without further insight, require an immense ammount of
calculation.  
  
Nevertheless, a complete G-structure classification of all
supersymmetric configurations in eleven dimensions (or in 
any other supergravity) will be of significant value. An in-depth
analysis of the constraints derived from a standard classification of
simple five dimensional supergravities has already yielded some very
interesting solutions, such as supersymmetric $AdS_5$ black holes
\cite{reall1}, \cite{reall2}, and supersymmetric black rings,
\cite{reall3}-\cite{gut2}. One would hope that a complete G-structure
analysis in eleven dimensions would reveal a wealth of surprising new
phenomena.

\section{Acknowledgements}
The author is supported by the National University of Ireland, EPSRC
and a Freyer studentship. Stimulating discussions with Marco Cariglia
during the early stages of this project are gratefully acknowledged.  

\appendix

\section{Conventions}
We work in almost plus signature, $\eta_{\m\v} = diag\, (-,+,\dots
,+)$. Seven dimensional spacetime indices are denoted by Greek letters
$\mu ,\nu, \dots$. Orthonormal indices on the four dimensional base
are denoted by lower case
Roman letters, $i$, $j$,...; spacetime indices on the base are denoted
by upper case Roman letters $M$, $N$,... 
In an orthonormal frame the Dirac algebra is 
\be
\lbrace \G_\mu , \G_ \nu \rbrace = 2 g_{\mu\nu}\,.
\ee
This tells us that $\G_0$ is antihermitian and the $\G_i$
$(i=1,\dots,6)$ are hermitian.  Following the appendix to Chapter 1 in
\cite{Salam:fm} we have that the charge conjugation matrix $C$ satisfies
\be
\label{con1}
C^T = C\,, \qquad C^\dagger C = \bfI \,,\qquad \G_\mu ^T = - C \G_\mu
C^{-1}\,. 
\ee
We can therefore choose 
\be
\label{con2}
C=\bfI \,.
\ee
This implies that $\G_0$ is real and the $\G_i$ are imaginary.  We
will choose a representation (there are two inequivalent ones) such
that
\be
\G_0 \G_1 \G_2 \G_3 \G_4 \G_5 \G_6 = - \bfI
\,. \label{eq:representation} 
\ee
We also have the identity
\be
\G_{\alpha_1 \dots \alpha_n} = \frac{(-)^{[n/2]+1}}{(7-n)!}
\e_{\alpha_1 \dots \alpha_n \beta_1 \dots \beta_{7-n}} \G^{\beta_1
  \dots \beta_{7-n}}\,. 
\ee
We choose the orientation to be given by $\epsilon^{0123456} =
+1$. Our null basis is defined by
\be
e^{\pm}=\frac{1}{\sqrt{2}}(e^0\pm e^6),
\end{equation}
so that $\e^{+-12345}=1$. The orientation on the base is chosen to be given by
$\e^{+-ijkl5}=\e^{ijkl}$.

The Dirac conjugate $\bar \e_a$  of an anticommuting
spinor $\e^a$ is defined as
\be
\bar \e_a  =(\e^a)^\dagger \G_0\,, 
\ee
and we also define 
\be 
\bar \e^a = \e^{ab}\,\bar \e_b\,,
\label{dirac}
\ee
where $\e^{ab}$ is a constant antisymmetric matrix satisfying
$\e_{ab}\, \e^{bc}= -\delta_a^c$ that is used to raise and lower spinor
indices according to $\e^a \equiv \e^{ab} \e_b$, and $\e^{12}=1$.
On the other hand the symplectic-Majorana conjugate $\e^C$ of $\e$ is
defined to be
\be
\label{majorana}
(\e^C)^a = (\e^T)_b\, .  
\ee
Symplectic-Majorana spinors are those
for which (\ref{dirac}) is equal to (\ref{majorana}), namely
\be
(\e^T)^a = \bar \e^a\,. \label{eq:Majorana_cond}
\ee
Given four spinors $\epsilon_1$, $\dots$, $\epsilon_4$, the Fierz
identity is 
\be 
    \overline{\epsilon_1}\epsilon_2 \overline{\epsilon_3}\epsilon_4 = 
    \frac{1}{8} 
    \left[ \overline{\e_1}\e_4 \overline{\e_3}\e_2 + 
      \overline{\e_1}\Gamma_\mu \e_4 \overline{\e_3} \Gamma^\mu \e_2 -
    \frac{1}{2} \overline{\e_1}\Gamma_{\mu\nu} \e_4 \overline{\e_3}
    \Gamma^{\mu\nu} \e_2 - \frac{1}{3!}
 \overline{\e_1}\Gamma_{\mu\nu\rho} \e_4 \overline{\e_3}
    \Gamma^{\mu\nu\rho} \e_2 \right] \label{eq:Fierz_id} . 
\ee

\section{Useful projections}
We give here the full set of projections satisfied by the fiducial 
basis spinor that we have employed in deriving the constraints.
\bea
\G^+\e&=&0,\\\G^{+-}\e&=&-\e,\\\G^{ijkl}\e&=&-\e^{ijkl}\e,\\A^{(-)}_{ij}\G^{ij}\e&=&0,\\\G^{ijk}\e&=&\e^{ijkl}\G_l\e,\\J^A_{ij}\G^{ij}\e^a&=&8iT^{Aa}_{\;\;\;\;\;b}\e^b,
 \\T^{Aa}_{\;\;\;\;\;b}\G_i\e^b&=&-\frac{i}{2}J^A_{ij}\G^j\e^a. 
\eea

\section{Spin connection components}
The exterior derivatives of the basis one forms are given by
\bea
de^+&=&e^+\wedge(\md\log
L+\pa_u\l)+e^5\wedge(-(LC)^{-1}\pa_z\l)\nonumber\\&+&e^+\wedge
e^5C^{-1}\pa_z\log L-L^{-1}\md\l,\\ de^-&=&e^+\wedge
L(\frac{1}{2}\md\mf+\frac{1}{2}\l\pa_u\mf-\pa_u\v+\s\pa_u\b)\nonumber\\&+&e^5\wedge
C^{-1}(-\frac{1}{2}\l\pa_z\mf+\pa_z\v-\s\pa_zB-\md
B)\nonumber\\&+&e^+\wedge
e^5LC^{-1}(\frac{1}{2}\pa_z\mf-\pa_uB)\nonumber\\&+&\md\v-\frac{1}{2}\md\mf\wedge\l-\md
B\wedge\s,\\de^5&=&e^+\wedge(-CL\pa_u\s)+e^5\wedge(\pa_z\s-\md\log
C)\nonumber\\&+&e^+\wedge e^5(-L\pa_u \log C)+C\md\s,\\de^i&=&\md
e^i+e^+\wedge(-L\pa_ue^i)+e^5\wedge(C^{-1}\pa_ze^i).
\eea   
Then the full set of nonzero spin connection components is
\bea
\o_{-+5}&=&\frac{1}{2C}\pa_z\log
L,\nonumber\\\o_{-+i}&=&\frac{1}{2}(\md\log
 L+\pa_u\l)_i,\nn\o_{-5i}&=&\frac{1}{2LC}(\pa_z\l)_i, \nn\o_{-ij}&=&
 \frac{1}{2L}\md\l_{ij},
\eea
\bea 
\o_{++5}&=&LC^{-1}(\frac{1}{2}\pa_z\mf-\pa_uB),\nn\o_{++i}&=&L(\frac{1}{2}
\md\mf+\frac{1}{2}\l
\pa_u\mf-pa_u\v+\s\pa_uB)_i,\nn\o_{+-5}&=&\o_{-+5},
\nn\o_{+-i}&=&\o_{-+i},\nn\o_{+5i}&=&
\frac{1}{2C}(-\frac{1}{2}\l\pa_z\mf+\pa_z\v-\s\pa_zB-\md 
B+LC^2\pa_u\s)_i,\nn\o_{+ij}&=&LM_{[ij]}+\frac{1}{2}(\md\v-\frac{1}{2}
\md\mf\wedge\l-\md 
  B\wedge\s)_{ij},
\eea
\bea
\o_{5+-}&=&\o_{-+5},\nn\o_{5+5}&=&L\pa_u\log
  C\nn\o_{5+i}&=&\o_{+5i},\nn\o_{5-i}&=&\o_{-5i},\nn\o_{55i}&=&(\md\log
  C-\pa_z\s)_i,\nn\o_{5ij}&=&-C^{-1}\Lambda_{[ij]}-\frac{1}{2}C\md\s_{ij},
\eea
\bea
\o_{i+-}&=&\o_{-+i},\nn\o_{i+5}&=&-\o_{+5i},\nn\o_{i+j}&=&LM_{(ij)}+\frac{1}{2}(\md\v-\frac{1}{2}\md\mf\wedge\l-\md B\wedge\s)_{ij},\nn\o_{i-5}&=&-\o_{-5i},\nn\o_{i-j}&=&\o_{-ij},\nn\o_{i5j}&=&-C^{-1}\Lambda_{(ij)}-\frac{1}{2}C\md\s_{ij},\nn\o_{ijk}&=&\hat{\o}_{ijk}+\s_i\Lambda_{[jk]}+\s_{[j}\Lambda_{k]i}+\s_{[j}\Lambda_{|i|k]}\nn&-&\l_iM_{[jk]}-\l_{[j}M_{k]i}-\l_{[j}M_{|i|k},
\eea
where $\hat{\o}_{ijk}$ is the spin connection on the base.

\section{Integrability Conditions \label{app:integrability}}
In \cite{us} it was shown that we may obtain the following
integrability condition from commuting the Killing spinor equation
with $\d\lambda$:
\bea
\sqrt{5}\G^{\m}[\mathcal{D}_{\m},\Delta_{\lambda}]\e^a&=&\Big(
\frac{1}{2}P+
\frac{1}{6}Q_{\m\v\s}\G^{\m\v\s}+\frac{e^{2\phi}}{96}d(e^{-2\phi}G)_{\m\v\s\t\rho}\G^{\m\v\s\t\rho}\Big)\e^a\nonumber\\&+&\Big(iR^A_{\m
  }\G^{\m}+\frac{ie^{-\phi}}{6}D(e^{\phi}F^{A})_{\m\v\s}\G^{\m\v\s}\Big)T^{Aa}_{\;\;\;\;\;b}\e^b\nonumber\\&+&\sqrt{5}\Big(\frac{1}{60}G_{\m\v\s\t}\G^{\m\v\s\t}\d^a_b+\frac{3i}{5}F^{A}_{\m\v }\G^{\m\v}T^{Aa}_{\;\;\;\;\;b}+(8h-g)\d^a_b\Big)\d\lambda^b,
\eea
where $D$ denotes the gauge-covariant exterior derivative, $P$, $Q$, $R$ are defined by
eqs.(\ref{eq:P},\ref{eq:Q},\ref{eq:R}), 
and the dilaton, four form and two form field equations are
respectively $P=0$, $Q=0$, $R^A=0$. It was also shown that the integrability condition for the Killing spinor
equation is
\bea
\G^{\v}[\mathcal{D}_{\m},\mathcal{D}_{\v}]\e^a&=&\Big[-\frac{1}{2}E_{\m\v}\G^{\v}+e^{2\phi}d(e^{-2\phi}G)^{\v\s\t\rho\xi}\Big(-\frac{1}{120}g_{\m\v}\G_{\s\t\rho\xi}\nn&+&\frac{1}{200}\G_{\m\v\s\t\rho\xi}\Big)+\frac{1}{10}Q^{\v\s\t}\Big(\frac{1}{2}\G_{\m\v\s\t}-g_{\m\v}\G_{\s\t}\Big)\Big]\e^a\nonumber\\&+&\Big[\frac{ie^{-\phi}}{5}D(e^{\phi}F^A)^{\v\s\t}\Big(2g_{\m\v}\G_{\s\t}+\frac{1}{6}\G_{\m\v\s\t}\Big)\nn&-&\frac{i}{5}R^{A\v}(-4g_{\m\v}+\G_{\m\v})\Big]T^{Aa}_{\;\;\;\;\;b}\e^b\nonumber\\&+&\Big[\pa_{\m}\phi\d^a_b-\frac{i}{25}F^{A\v\s}(8g_{\m\v}\G_{\s}-\G_{\m\v\s})T^{Aa}_{\;\;\;\;\;b}+\frac{2}{5}m^{\prime}\G_{\m}\d^a_b\nonumber\\&+&\frac{1}{25}G^{\v\s\t\rho}\Big(-\frac{2}{3}g_{\m\v}\G_{\s\t\rho}+\frac{1}{4}\G_{\m\v\s\t\rho}\Big)\d^a_b
  \Big]\sqrt{5}\d\lambda^b=0,
\eea
where the Einstein equations are $E_{\m\v}=0$.  

When a single null Killing spinor exists one may readily verify that is is sufficient to
impose the Bianchi identities for the forms, the field equations for
$G_{(4)}$, the $+$ component of the Yang-Mills field equations and the
$++$ component of the Einstein equations. When additional Killing
spinors are incorporated, more of the field equations and Bianchi
identities are identically satisfied; working out which ones may be
done straightforwardly case by case.


\begin{thebibliography}{99}
\bibitem{tod} K. Tod, {\it All metrics admitting supercovariantly
  constant spinors}, Phys. Lett. B 121 241 (1983)

\bibitem{gaunt} J. Gauntlett, J. Gutowski, C. Hull, S. Pakis and
  H. Reall, {\it All supersymmetric solutions of minimal supergravity
  in five dimensions}, Class. Quant. Grav. 20 (2003) 4587

\bibitem{gaunt2} J. Gauntlett and J. Gutowski, {\it All supersymmetric
  solutions of minimal gauged supergravity in five dimensions},
  Phys. Rev. D68, 105009 (2003)

\bibitem{J1} D. Martelli and J. Sparks, {\it G-Structures, fluxes and
  calibrations in M-theory}, Phys. Rev. D68 085014 (2003)

\bibitem{reall} J. Gutowski, D. Martelli and H. Reall, {\it All
  supersymmetric solutions of minimal supergravity in six dimensions},
  hep-th/0306235.

\bibitem{us1} M. Cariglia and O. A. P. Mac Conamhna, {\it The
  general form of supersymmetric solutions of N=(1,0) U(1) and SU(2)
  gauged supergravities in six dimensions}, Class.Quant.Grav 21 (2004)
  3171, hep-th/0402055.


\bibitem{Waldram} J. P. Gauntlett, D. Martelli, D. Waldram, {\it
  Superstrings with Intrinsic Torsion}, hep-th/0302158. 

\bibitem{klemm} M. Caldarelli and D. Klemm, {\it All supersymmetric
  solutions of N = 2, D = 4 gauged supergravity}, JHEP 0309 019 (2003)



 \bibitem{gaunt1} J. Gauntlett and S. Pakis, {\it The geometry of D =
  11 Killing spinors}, JHEP 0304 039 (2003)

\bibitem{gaunt3} J. Gauntlett, J. Gutowski and S. Pakis, {\it The
  geometry of D = 11 null Killing spinors}, hep-th/0311112



\bibitem{gmpw}  J. P. Gauntlett, D. Martelli, S. Pakis, and
  D. Waldram, {\it  G-Structures and Wrapped NS5-Branes}, hep-th/0205050.



\bibitem{Martelli_Sparks} 
 J. P. Gauntlett, D. Martelli, J. Sparks and D. Waldram, {\it
 Supersymmetric $AdS_5$ solutions of M-theory}, hep-th/0402153.  

\bibitem{klemmn} S. L. Cacciatori, M. M. Caldarelli, D. Klemm and
  D. S. Mansi, {\it More on BPS solutions of N=2, D=4 gauged
  supergravity}, hep-th/0406238.

\bibitem{andre}  A. Lukas and P.M. Saffin, {\it M-theory
  compactification, fluxes and $AdS_4$}, hep-th/0403235.

\bibitem{gianguido}  G. Dall'Agata, {\it On supersymmetric solutions
  of type IIB supergravity with general fluxes}, hep-th/0403220.

\bibitem{hackett-jones}  E. J. Hackett-Jones and D. J. Smith, {\it
  Type IIB Killing spinors and calibrations}, hep-th/0405098.

\bibitem{frey} A. R. Frey, {\it Notes on SU(3) Structures in Type IIB
  Supergravity}, hep-th/0404107.

\bibitem{caccia} S. L. Cacciatori, M. M. Caldarelli, D. Klemm and
  D. S. Mansi, {\it More on BPS solutions of N=2, D=4 gauged
  supergravity}, hep-th/0406238.

\bibitem{us} M. Cariglia and O. Mac Conamhna, {\it Timelike Killing
  spinors in seven dimensions}, hep-th/0407127.

\bibitem{minasian} M. Grana, R. Minasian, M. Petrini and
  A. Tomasiello, {\it Supersymmetric backgrounds from generalised
  Calabi-Yau manifolds}, hep-th/0406137.

\bibitem{duff} M. J. Duff and J. T. Liu, {\it Hidden spacetime
  symmetries and generalised holonomy in M-theory}, Nucl.Phys. B674
  (2003) 217-230, hep-th/0303140.

\bibitem{hull} C. Hull, {\it Holonomy and symmetry in M-theory},
  hep-th/0305039.

\bibitem{tsimpis} G. Papadopoulos and D. Tsimpis, {\it The holonomy of
  the supercovariant connection and Killing spinors}, JHEP 0307 (2003)
  018, hep-th/0306117. 

 \bibitem{Townsend:1983kk}
P.~K.~Townsend and P.~van Nieuwenhuizen, {\it Gauged Seven-Dimensional
  Supergravity}, 
Phys.\ Lett.\  {\bf B125} (1983) 41.


\bibitem{Mezincescu:ta}
L.~Mezincescu, P.~K.~Townsend and P.~van Nieuwenhuizen,
{\it Stability Of Gauged D = 7 Supergravity And The Definition Of
  Masslessness In (Ads) In Seven-Dimensions}, 
Phys.\ Lett.\ {\bf B143} (1984) 384.

\bibitem{chamseddine} A. H. Chamseddine and W.A. Sabra, {\it  D = 7
  SU(2) gauged supergravity from D=10 supergravity},
  Phys.Lett.B476:415-419,2000, hep-th/9911180.

\bibitem{pope} H. L\"{u} and C. N. Pope, {\it Exact embedding of $N=1$,
  $D=7$ gauged supergravity in $D=11$}, Phys.Lett. B467 (1999) 67-72,
  hep-th/9906168. 

\bibitem{Pilch} P. K. Townsend, K. Pilch and P. Van Nieuwenhuizen,
  {\it Self-duality in odd dimensions}, Phys. Lett. {\bf B136} (1984)
  38, Addendum-ibid {\bf B137} (1984) 443.

\bibitem{bryant} R. L. Bryant, {\it Pseudo-Riemannian metrics with
  parallel spinor fields and vanishing Ricci tensor}, math.DG 0004073.


\bibitem{wrap} J. P. Gauntlett, N. Kim and D. Waldram, {\it
 M-Fivebranes Wrapped on Supersymmetric Cycles}, Phys.Rev. D63 (2001)
 126001, hep-th/0012195.

\bibitem{ach} B. S. Acharya, J. P. Gauntlett and N. Kim, {\it
  Fivebranes wrapped on associative three-cycles}, Phys.Rev. D63
  (2001) 106003, hep-th/0011190. 

\bibitem{carlos} J. Maldacena and C. Nunez, {\it Supergravity
  description of field theories on curved manufolds and a no go
  theorem}, Int.J.Mod.Phys. A16 (2001) 822, hep-th/007018.

\bibitem{Salam:fm}
A.~Salam and E.~Sezgin,
``Supergravities In Diverse Dimensions. Vol. 1, 2.''

\bibitem{jose} J. M. Figueroa-O'Farrill, {\it Breaking the M-waves},
Class.Quant.Grav. 17 (2000) 2925, hep-th/9904124.

\bibitem{reall1} J. B. Gutowski and H. S. Reall, {\it Supersymmetric AdS5
  black holes}, JHEP 0402 (2004) 006, hep-th/0401042.

\bibitem{reall2} J. B. Gutowski and H. S. Reall, {\it General
  supersymmetric AdS5 black holes}, JHEP 0404 (2004) 048,
  hep-th/0401129.

\bibitem{reall3} H. Elvang, R. Emparan, D. Mateos and H. S. Reall,
  {\it A supersymmetric black ring}, hep-th/0407065.

\bibitem{gut1} J. P. Gauntlett and J. B. Gutowski, {\it Concentric
  black rings}, hep-th/0408010.

\bibitem{gut2} J. P. Gauntlett and J. B. Gutowski, {\it General
  Concentric Black Rings},  hep-th/0408122. 

\end{thebibliography}
\end{document}